\documentclass{iopart}
\usepackage{euscript,iopams,amssymb,amsfonts,graphicx,bm}
\usepackage{pgfplots,setstack}

\usepackage{float}
\usepackage{epsfig}
\usepackage{esint}
 \usepackage{tikz-cd}

\usepackage{bm,braket}

\eqnobysec

\newcommand{\dis}{\displaystyle}

\renewcommand{\theequation}{\arabic{section}.\arabic{equation}}

\newcommand{\hcalQ}{\widehat{\mathcal Q}}
\newcommand{\calP}{{\mathcal P}}
\newcommand{\calQ}{{\mathcal Q}}
\newcommand{\calF}{{\mathcal F}}
\newcommand{\calJ}{{\mathcal J}}
\newcommand{\R}{{\mathbb R}}

\newcommand{\X}{\mathbf{X}}
\renewcommand{\P}{\mathbb{P}}

\newcommand{\PP}{\widetilde{\mathcal P}}
\newcommand{\QQ}{\widetilde{\mathcal Q}}
\newcommand{\x}{\mathbf{x}}
\renewcommand{\a}{\mathbf{a}}
\newcommand{\A}{\mathbf{A}}

\renewcommand{\e}{{\mathrm e}}
\newcommand{\E}{{\mathbb E}}

\newcommand{\n}{\mathbf n}
\newcommand{\z}{\mathbf z}
\newcommand{\calT}{{\mathcal T}}
\renewcommand{\P}{\mathbb P}
\newcommand{\p}{\widetilde{p}}

\renewcommand{\j}{\widetilde{j}}

\newcommand{\ellh}{\hat{\ell}}

\begin{document}

 \title[Encounter-based model of a run-and-tumble particle]{Encounter-based model of a run-and-tumble particle II: absorption at sticky boundaries}

\author{Paul C. Bressloff}
\address{Department of Mathematics, University of Utah 155 South 1400 East, Salt Lake City, UT 84112}

\begin{abstract} 
In this paper we develop an encounter-based model of a run-and-tumble particle (RTP) confined to a finite interval $[0,L]$ with partially absorbing, sticky boundaries at both ends. We assume that the particle switches between two constant velocity states $\pm v$ at a rate $\alpha$. Whenever the particle hits a boundary, it becomes stuck by pushing on the boundary until either a tumble event reverses the swimming direction or it is permanently absorbed. We formulate the absorption process by identifying the first passage time (FPT) for absorption with the event that the time $A(t)$ spent attached to either wall up to time $t$ (the occupation time) crosses some random threshold $\widehat{A}$. Taking $\Psi(a)\equiv \P[\widehat{A}>a]$ to be an exponential distribution, $\Psi[a]=\e^{-\kappa_0 a}$, we show that the joint probability density for particle position $X(t)$ and velocity state $\sigma(t)=\pm $ satisfies a well-defined boundary value problem (BVP) with $\kappa_0$ representing a constant absorption rate. The solution of this BVP determines the so-called occupation time propagator, which is the joint probability density for the triplet $(X(t),A(t),\sigma(t))$. The propagator is then used to incorporate more general models of absorption based on non-exponential (non-Markovian) distributions $\Psi(a)$. We illustrate the theory by calculating the mean FPT (MFPT) and splitting probabilities for absorption. We also show how our previous results for partially absorbing, non-sticky boundaries can be recovered in an appropriate limit. Absorption now depends on the number of collisions $\ell(t)$ of the RTP with the boundary. Finally, we extend the theory by taking absorption to depend on the individual occupation times at the two ends.

\end{abstract}

\maketitle

\newpage
\section{Introduction}

In a previous paper, we developed an encounter-based model of  a run-and-tumble particle (RTP) confined to a one-dimensional (1D) domain with a partially absorbing boundary \cite{Bressloff22rtp}. An RTP is one of the simplest examples of active matter, in which a particle randomly switches between a left-moving ($\sigma=-$) and a right-moving ($\sigma=+)$ constant velocity state of speed $v$ at some Poisson rate $\alpha$. It is inspired by the observed `run-and-tumble' motion of bacteria such as {\em E. coli} \cite{Berg04}. In the presence of an attractive chemotactic signal, the switching rate becomes biased so that runs towards the source of the signal tend to persist. Mathematically speaking, the position $X(t)$ of an RTP evolves according to a two-state velocity jump process. Velocity jump processes arise in a wide range of other applications in cell biology. For example, $X(t)$ could represent the position of a motor-cargo complex undergoing active transport along a microtubule \cite{Newby10,Bressloff13}. Alternatively, $X(t)$ could denote the position of the tip of a polymer filament undergoing successive rounds of growth and shrinkage. A classical example is the Dogterom-Leibler model of microtubular catastrophes \cite{Dogterom93}, which play an important role in cell mitosis. A more recent example is cytoneme-based search and capture during embryogenesis \cite{Bressloff19}.  

There is a rapidly growing literature on the stochastic dynamics of RTPs at the single particle level, covering topics such as the probability density of freely moving RTPs \cite{Martens12,Gradenigo19,Singh19} and RTPs confined by a potential \cite{Dhar19,Sevilla19,Dor19}, RTPs under stochastic resetting \cite{Evans18,Bressloff20,Santra20a} and the analysis of first-passage times (FPTs) \cite{Angelani14,Angelani15,Angelani17,Malakar18,Demaerel18,Doussal19}. 
Our previous paper focused on the FPT problem for an RTP confined to an interval $[0,L]$ with a partially absorbing boundary at one end. The particular case of a constant rate of absorption (reactivity) was solved by Angelani \cite{Angelani14,Angelani15}. We considered a more general class of partially absorbing boundaries by adapting the encounter-based model of diffusion-mediated surface absorption \cite{Grebenkov20,Grebenkov22,Bressloff22,Bressloff22b}. That is, we assumed that the reactivity is a function of the number of collisions (discrete local time) $\ell(t)$ between the RTP and the boundary over the time interval $[0,t]$.\footnote{In the case of a Brownian particle, $\ell(t)$ is a Brownian functional known as the boundary local time, which is a nondecreasing continuous function of time $t$ \cite{McKean75,Majumdar05}.} We then introduced the joint probability density or generalized propagator $P_{\sigma}(x,\ell,t)$ for the triplet $(X(t),\ell(t),\sigma(t))$, $\sigma(t)=\pm$, in the case of a perfectly reflecting boundary. Partial absorption was incorporated into the model
 by imposing the stopping time 
${\mathcal T}=\inf\{t>0:\ \ell(t) >\widehat{\ell}\}$,
 with $\widehat{\ell}$ a random local time threshold. Given the probability distribution $\Psi(\ell) = \P[\ellh>\ell]$ with $\Psi(0)=1$, the marginal probability densities for particle position were defined according to
 \begin{equation}
 \label{pc}
 p_{\sigma}^{\Psi}(x,t)=\sum_{\ell=0}^{\infty} \Psi(\ell-1)P_{\sigma}(x,\ell,t)d\ell,
 \end{equation}
 with $\Psi(-1)\equiv 0$.
 We then showed that the discrete Laplace transform $\widetilde{P}_{\sigma}(x,z,t)=\sum_{\ell=0}^{\infty} z^{\ell} P_{\sigma}(x,\ell,t)$, which is equivalent to taking a geometric distribution $\Psi(\ell)=z^{\ell}$, satisfied the boundary value problem (BVP) considered in Ref. \cite{Angelani15} for a constant rate of absorption $\kappa_0$, with $z=1/(1+\kappa_0/v)$. Hence, a general non-Markovian form of absorption can be modeled by solving the propagator BVP for a constant absorption rate $\kappa_0$, setting $\kappa_0=v(1-z)/z$, inverting the discrete Laplace transform with respect to $z$, and then calculating the marginal probability density for a general distribution $\Psi(\ell)$:
 \begin{equation}
  p_{\sigma }^{\Psi}(x,t)=\sum_{\ell=0}^{\infty} \Psi(\ell-1){\mathcal L}_{\ell}^{-1}\widetilde{P}_{\sigma}(x,z,t),
  \label{pc2}
  \end{equation}
  where ${\mathcal L}^{-1}_{\ell}$ denotes the inverse discrete Laplace transform.    

Within the context of active matter, RTP models provide an analytically tractable framework for investigating self-organizing phenomena such as the tendency of active particles to accumulate at walls, which can occur even if inter-particle interactions are ignored (see the review \cite{Bechinger16} and references therein). An accumulation mechanism may be incorporated into a model of confined RTP motion using a so-called sticky boundary condition: whenever the particle hits a hard wall, it becomes stuck by pushing on the boundary until a tumble event reverses the swimming direction. A 1D version of such a model was analyzed in Ref. \cite{Angelani17}, both for totally reflecting and partially absorbing walls. In addition, an effective attractive or repulsive force at the wall was included by taking the tumbling rate of particles at the wall to differ from the tumbling rate in the bulk. Sticky boundaries also arise within the context of the growth and shrinkage of polymer filaments \cite{Zelinski12,Mulder12,Bressloff19}. For example, a nucleation site for polymer formation can be modeled as a sticky boundary.

\begin{figure}[b!]
\centering
\includegraphics[width=13cm]{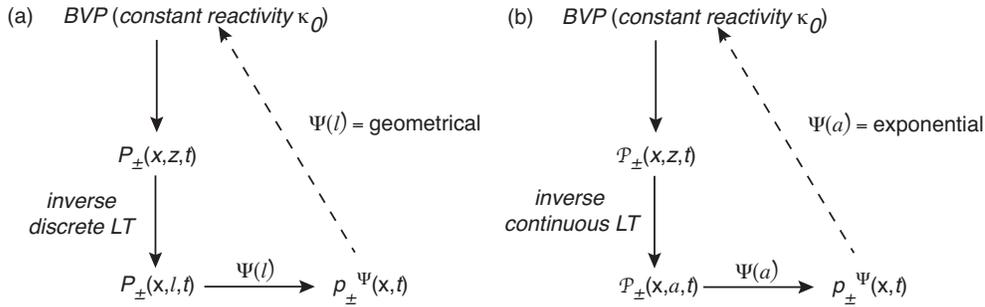}
\caption{Diagrams illustrating the encounter-based framework for an RTP in a domain with a partially absorbing boundary. (a) Non-sticky boundary. The solution of the BVP in the case of a constant reactivity $\kappa_0$ generates the discrete Laplace transform $\widetilde{P}_{\pm}(x,z,t)$ of the discrete local time propagator $P_{\pm}(x,\ell,t)$ with $z=1/(1+\kappa_0/v)$, and $p_{\pm}^{\Psi}(\x,t)$ is given by equation (\ref{pc2}). (b) Sticky boundary. The solution of the BVP in the case of a constant reactivity $\kappa_0$ generates the Laplace transform $\widetilde{\calP}_{\pm}(x,z,t)$ of the continuous occupation time propagator $\calP_{\pm}(x,a,t)$ with $z=\kappa_0$, and $p_{\pm}^{\Psi}(\x,t)$ is given by equation (\ref{pca2})}
\label{fig1}
\end{figure}

In this paper we extend our previous work on encounter-based models of RTP-based absorption  \cite{Bressloff22rtp} to the case of sticky boundaries at both ends of an interval $[0,L]$.   
The resulting absorption mechanism differs significantly from the non-sticky case, since absorption now occurs when the particle is in a bound state, that is, attached to a wall. Hence, rather than formulating absorption in terms of a discrete local time that counts the number of reflecting collisions with the walls, it is more natural to consider the occupation time $A(t)$, which is the amount of time that the RTP spends in a bound state over the time interval $[0,t]$. (Occupation times also arise in the case of Brownian motion in an interval with a partially absorbing interior \cite{Bressloff22,Bressloff22a}.) We proceed by defining an occupation time propagator $\calP_{\sigma}(\x,a,t)$ for the triplet $(\X(t),a(t),\sigma(t))$, introducing a random occupation time threshold $\widehat{A}$ with distribution $\Psi(a)$ and setting
 \begin{equation}
 \label{pca}
 p_{\sigma}^{\Psi}(\x,t)=\int_0^{\infty} \Psi(a)\calP_{\sigma}(\x,a,t)da.
 \end{equation}
 We show that Laplace transforming the propagator with respect to $a$, which is equivalent to taking an exponential distribution $\Psi(a)=\e^{-za}$, leads to the boundary value problem (BVP) for a constant rate of absorption $\kappa_0=z$ that was analyzed in Ref. \cite{Angelani17}. Hence, we incorporate a more general form of absorption at a sticky boundary by solving the propagator BVP for a constant absorption rate $\kappa_0$, setting $\kappa_0=z$, inverting the Laplace transform with respect to $z$, and then calculating the marginal probability density for a general distribution $\Psi(\ell)$ according to 
 \begin{equation}
 \label{pca2}
 p_{\sigma}^{\Psi}(\x,t)=\int_0^{\infty} \Psi(a){\mathcal L}^{-1}_a\widetilde{\calP}_{\sigma}(\x,z,t)da.
 \end{equation}
 A summary of the encounter-based formulations of RTP absorption at a non-sticky and sticky boundary, respectively, is shown in Fig. \ref{fig1}.
 
 The structure of the paper is as follows. In \S 2 we introduce the basic model of RTP confinement in an interval with sticky boundaries at both ends. We calculate the MFPT and splitting probabilities in the case of a constant absorption rate, and show how previous results are obtained in various limits. In \S 3 we formulate the encounter-based version of the model, and describe how to recover the discrete local time formulation of non-sticky boundary conditions. In \S 4 we calculate the corresponding MFPT and splitting probabilities for generalized absorption, and explore how these quantities depend on the stickiness parameter $\gamma$. Finally, in \S 5, we extend the theory by taking absorption to depend on the individual occupation times $A_0(t)$ and $A_L(t)$ at the ends $x=0$ and $x=L$, respectively. The analysis is more involved, since it is necessary to consider a propagator that depends on a pair of occupation times $A_0(t)$ and $A_L(t)$, and to formulate absorption in terms of a corresponding pair of independent, random thresholds $\widehat{A}_0$ and $\widehat{A}_L$. Performing a double Laplace transform of the propagator with respect to these occupation times leads to a BVP with the constant rate of absorption at each end being distinct, which is straightforward to solve. However, in order to determine the marginal probability density $ p_{\sigma}^{\Psi}(\x,t)$, we have to calculate a double inverse Laplace transform and then integrate the resulting expression with respect to a pair of occupation time thresholds weighted by their corresponding probability distributions:
 \begin{equation}
 \label{pca3}
 p_{\sigma}^{\Psi}(\x,t)=\int_0^{\infty}da\, \Psi_0(a)\int_0^{\infty}da'\Psi_L(a'){\mathcal L}^{-1}_a{\mathcal L}^{-1}_{a'}\widetilde{\calP}_{\sigma}(\x,z_0,z,L,t).
 \end{equation}

\setcounter{equation}{0}
\section{Run-and-tumble particle in an interval with sticky boundaries}

\begin{figure}[t!]
\centering
\includegraphics[height=4cm]{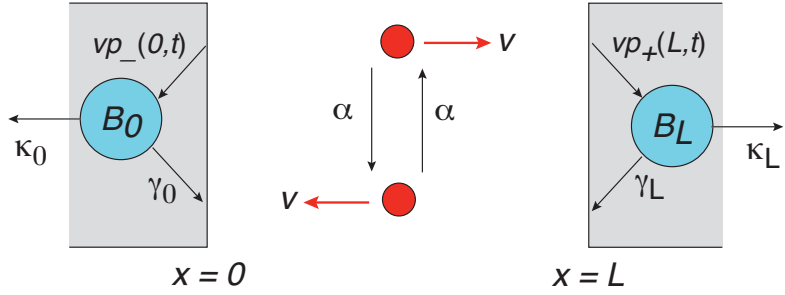}
\caption{Schematic representation of an RTP confined to the domain $[0,L]$ with sticky boundaries at both ends. The particle attaches to the boundary at $x=0,L$ by entering the bound state $B_{0,L}$, and subsequently re-enters the bulk domain at a rate $\gamma_{0,L}$. Whilst in a bound state, the particle can be absorbed at a rate $\kappa_{0,L}$. There a number of possible interpretations of the 1D model. (i) The particle represents the position $X(t)$ of a bacterium confined to an infinitely extended channel of width $L$, with the ends corresponding to the channel walls \cite{Berg04}. An absorption event could be the death of the cell.  (ii) $X(t)$ represents the tip of a growing and shrinking polymer filament that nucleates from a source cell at the end $x=0$, say, and temporarily binds to a target cell at $x=L$ \cite{Bressloff19}. Absorption could correspond to destruction of the nucleation site. (iii) $X(t)$ represents the position of a molecular motor on a polymer filament delivering cargo from the cell body of a neuron to a location along the axon \cite{Newby10,Bressloff13}. Absorption could correspond to removal of the motor from the filament for possible recycling.}
\label{fig2}
\end{figure}

A schematic illustration of the basic 1D model is shown in Fig. \ref{fig2}, and a number of possible physical interpretations are listed in the figure caption. The model considers an RTP confined to the interval $x\in [0,L]$ that randomly switches between two constant velocity states labeled by $\sigma=\pm $ with $v_{\sigma}=v\delta_{\sigma,+}-v\delta_{\sigma,-}$ and $v>0$.  Furthermore, suppose that the particle reverses direction (switches between the velocity states) according to a Poisson process with rate $\alpha$. The position $X(t)$ of the particle at time $t$ evolves according to the piecewise deterministic equation
\begin{equation}
\label{PDMP}
\frac{dX}{dt}=v_{\sigma(t)},
 \end{equation}
where $\sigma(t)$ is a dichotomous noise process. Let $p_{\sigma}(x,t)$ be the probability density that at time $t$ the end of the MT is at $X(t)=x$ and in the discrete state $\sigma(t)=\sigma=\pm$. The associated evolution equation is given by
\numparts
\begin{eqnarray}
\label{DLa}
\frac{\partial p_{+}}{\partial t}&=-v \frac{\partial p_{+}}{\partial x}-\alpha p_{+}+\alpha p_{-},\\
\frac{\partial p_{-}}{\partial t}&=v \frac{\partial p_{-}}{\partial x}-\alpha p_{-}+\alpha p_{+}.
\label{DLb}
\end{eqnarray}
\endnumparts
This is supplemented by the initial conditions $X(0)=x_0$ and $\sigma(0)=\sigma $ with probability $\rho_{\sigma} =1/2$. Note that under the change of variables
\begin{equation}
\label{cov}
 p(x,t )=p_+(x,t )+p_-(x,t ),\quad j(x,t )=v [p_+(x,t )-p_-(x,t )],
\end{equation}
with $p$ the marginal probability for particle position and $j$ the probability flux, we can rewrite equations (\ref{DLa}) and (\ref{DLb}) in the form
\numparts
\begin{eqnarray}
\label{JDLa}
\frac{\partial p}{\partial t}&=-\frac{\partial j}{\partial x},\\
\frac{\partial j}{\partial t}&= -v^2\frac{\partial p}{\partial x}-2\alpha j.
\label{JDLb}
\end{eqnarray}
\endnumparts
Equations (\ref{DLa}) and (\ref{DLb}) represent the differential Chapman-Kolmogorov (CK) for a symmetric two-state velocity jump process. 

Following \cite{Angelani17}, we will take the ends $x=0,L$ to be partially absorbing, sticky boundaries. That is, whenever the particle hits a boundary, its velocity immediately drops to zero and it sticks to the boundary until it reverses its velocity state at a Poisson rate $\gamma_0$ and $\gamma_L$ at the ends $x=0$ and $x=L$, respectively.  Whilst it is attached to the boundary, it can be absorbed or killed at the corresponding rates $\kappa_0$ and $\kappa_L$.
Mathematically speaking, one can incorporate a sticky boundary at $x=0$ by introducing the probability $Q_0(t)$ that at time $t$ the particle is attached to the left-hand boundary in the bound state $B_0$. The boundary condition at $x=0$ takes the form
\numparts
\begin{equation}
\label{mstick0}
\gamma_0 Q_0(t)=vp_+(0,t)=\frac{vp(0,t)+j(0,t)}{2},
\end{equation}
with $Q_0(t)$ evolving according to the equation
\begin{equation}
\label{stick0}
\frac{dQ_0}{dt}=vp_-(0,t)-(\gamma_0+\kappa_0) Q_0(t)=-j(0,t)-\kappa_0 Q_0(t).
\end{equation}
\endnumparts
Similarly, introducing the probability $Q_L(t)$ that at time $t$ the particle is stuck at the right-hand boundary in the bound state $B_L$, we have
\numparts 
\begin{equation}
\label{mstick1}
\gamma_L Q_L(t)=vp_-(L,t)=\frac{vp(L,t)-j(L,t)}{2},
\end{equation}
and
\begin{equation}
\label{stick1}
\frac{dQ_L}{dt}=vp_+(L,t)-(\gamma_L+\kappa_L) Q_L(t)=j(L,t)-\kappa_LQ_L(t).
\end{equation}
\endnumparts
Combining the two bound states $B_{0,L}$ with the two velocity states $\sigma=\pm$, we have a total of four states. The internal state at time $t$ is thus $\sigma(t)\in \{+,-,B_0,B_L\}$.

 It is instructive to consider various limiting cases. First, if $\kappa_0=0=\kappa_L$ then the sticky boundaries are non-absorbing and we have the conservation equation
\begin{equation}
\label{norm}
\int_0^{L}p(x,t)dx+Q_0(t)+Q_L(t)=1.
\end{equation}
The steady-state probability density $p_{\infty}=\lim_{t\rightarrow \infty} p(x,t)$ is spatially uniform with $Lp_{\infty}+2Q_{\infty}=1$ where $p_{\infty}=\lim_{t\rightarrow \infty} p(x,t)$. Second, if $\gamma_0=0$, say, then the boundary $x=0$ is totally absorbing in the sense that, when the RTP hits the boundary, it cannot reenter the bulk domain. However, the RTP still exists in a bound state until it is finally killed at the rate $\kappa_0$. A more subtle limit is the one that recovers the non-sticky, partially reflecting boundary condition considered in our previous paper \cite{Bressloff22rtp}. This involves taking the double limit $\kappa_0\rightarrow \infty$ and $\gamma_0 \rightarrow \infty$ such that the ratio $\kappa_0/\gamma_0 $ is fixed. Defining
\begin{equation}
\label{kay0}
k_0=\lim_{\kappa_0\rightarrow \infty}\lim_{\gamma\rightarrow \infty} \frac{\kappa_0v}{\gamma},
\end{equation}
we obtain the RTP equivalent of a Robin boundary condition \cite{Angelani15}, $j(0,t)=-k_0p_+(0,t)$. Applying an analogous double limit at $x=L$ would lead to the boundary condition $j(L,t)=k_Lp_-(L,t)$.

\subsection{Solution in Laplace space}

In preparation for the encounter-based formulation developed in subsequent sections, it is useful to perform the analysis of  the RTP model, following along similar lines to Ref. \cite{Angelani17}. (Only the details of the case $\kappa_0=0=\kappa_L$ were presented in Ref. \cite{Angelani17}.) For simplicity,  we take $\gamma_0=\gamma_L=\gamma$ and $\kappa_0=\kappa_L =\kappa$. We will relax the latter constraint in \S 5. First,
Laplace transforming equations (\ref{JDLa}) and (\ref{JDLb}) with $\widetilde{p}(x,s)=\int_0^{\infty}\e^{-st}p(x,t)dt$ etc., we have
\[s\widetilde{p}-\delta(x-x_0)=-\frac{\partial \widetilde{j}}{\partial x},\quad s\widetilde{j}=-v^2\frac{\partial \widetilde{p}}{\partial x}-2\alpha \widetilde{j},\]
which can be rearranged to give
\begin{equation}
\label{stickyL}
D(s)\frac{\partial^2 \widetilde{p}}{\partial x^2}-s \widetilde{p}=- \delta(x-x_0), \quad D(s)= \frac{v^2}{s+2\alpha}.
\end{equation}
Laplace transforming the boundary equations (\ref{stick0}) and (\ref{stick1}) gives
\begin{equation}
\fl v\widetilde{p}(0,s)=-\frac{2\gamma+s+\kappa}{s+\kappa}\widetilde{j}(0,s),\quad v\widetilde{p}(L,s)=\frac{2\gamma+s+\kappa}{s+\kappa}\widetilde{j}(L,s),
\label{stickyL2}
\end{equation}
which, on expressing $\widetilde{j}$ in terms of $\partial_x\widetilde{p}$, yields 
\begin{equation}
\label{sticky2}
 \left  . \frac{\partial \widetilde{p}(x,s)}{\partial x}\right |_{x=0}=\Gamma(s)\widetilde{p}(0,s),\quad \left  .  \frac{\partial \widetilde{p}(x,s)}{\partial x}\right |_{x=L}=-\Gamma(s)\widetilde{p}(L,s),
\end{equation}
with
\begin{equation}
\Gamma(s)=\frac{s+\kappa}{v}\frac{s+2\alpha}{s+\kappa+2\gamma}.
\end{equation}

A solution of equation (\ref{stickyL}) for $0\leq x<x_0$ that satisfies the boundary condition at $x=0$ takes the form
\numparts
\begin{equation}
\label{pl}
\fl \widetilde{p}_<(x,s)= \e^{\sqrt{s/D(s)}x}+\Lambda(s)\e^{-\sqrt{s/D(s)}x},\quad \Lambda(s)=\frac{\sqrt{s/D(s)}-\Gamma(s)}{\sqrt{s/D(s)}+\Gamma(s)}.\end{equation}
Similarly, a solution of equation (\ref{stickyL}) for $L>x>x_0$ that satisfies the boundary condition at $x=L$ is
\begin{equation}
\label{pl}
\widetilde{p}_>(x,s)= \e^{\sqrt{s/D(s)}(L-x)}+\Lambda(s)\e^{-\sqrt{s/D(s)}(L-x)}.\end{equation}
\endnumparts
Imposing continuity of the solution across $x=x_0$ and the flux discontinuity condition $D(s)\partial_x\p(x_0^+,s)-D(s)\partial_x\p(x_0^-,s)=-1$ implies that
\begin{equation}
\p(x,s)=\left \{ \begin{array}{cc} A(s)\p_<(x,s)\p_>(x_0,s),& 0\leq x < x_0 \\
A(s)\p_<(x_0,s)\p_>(x,s),& x< x_0 \leq L, \end{array} \right .
\label{psticky}
\end{equation}
with
\begin{equation}
A(s)=\frac{1}{2\sqrt{sD(s)}}\frac{1}{ \e^{\sqrt{s/D(s)}L}-\Lambda(s)^2\e^{-\sqrt{s/D(s)}L} }.
\end{equation}

We can now determine the Laplace transforms of the probabilities in the bound states:
\numparts
\begin{eqnarray}
\fl   \widetilde{Q}_0(s)&=\frac{v\p(0,s)+\widetilde{j}(0,s)}{2\gamma}\nonumber\\
\fl &=\frac{A(s)}{2\gamma }\p_>(x_0,s)\left [v (1+\Lambda(s))-\sqrt{sD(s)}(1-\Lambda(s) \right ]\nonumber \\
\fl &=\frac{1}{2\gamma }\frac{v /D(s) - \Gamma(s)}{\sqrt{s/D(s)}+\Gamma(s)}\frac{e^{\sqrt{s/D(s)}(L-x_0)}+\Lambda(s)\e^{-\sqrt{s/D(s)}(L-x_0)}}{ \e^{\sqrt{s/D(s)}L}-\Lambda(s)^2\e^{-\sqrt{s/D(s)}L}}
\nonumber \\
\fl &= \frac{s+2\alpha}{s+\kappa+2\gamma}\frac{1/v}{\sqrt{s/D(s)}+\Gamma(s)}\frac{e^{\sqrt{s/D(s)}(L-x_0)}+\Lambda(s)\e^{-\sqrt{s/D(s)}(L-x_0)}}{ \e^{\sqrt{s/D(s)}L}-\Lambda(s)^2\e^{-\sqrt{s/D(s)}L}},
\label{LTQ0}
\end{eqnarray}
and
\begin{eqnarray}
\fl   \widetilde{Q}_L(s)&=\frac{v\p(L,s)-\widetilde{j}(L,s)}{2\gamma}\nonumber\\
\fl &=\frac{A(s)}{2\gamma }\p_<(x_0,s)\left [v (1+\Lambda(s))-\sqrt{sD(s)}(1-\Lambda(s) \right ]\nonumber \\
\fl &= \frac{s+2\alpha}{s+\kappa+2\gamma}\frac{1/v}{\sqrt{s/D(s)}+\Gamma(s)}\frac{e^{\sqrt{s/D(s)} x_0}+\Lambda(s)\e^{-\sqrt{s/D(s)}x_0}}{ \e^{\sqrt{s/D(s)}L}-\Lambda(s)^2\e^{-\sqrt{s/D(s)}L}}.
\label{LTQL}
\end{eqnarray}
\endnumparts
A number of features are worth highlighting.
\medskip

\noindent (i) If $x_0=L/2$ so that the particle starts in the middle of the domain, then we recover the result quoted in Ref. \cite{Angelani17} (after defining $c(s)=\sqrt{s/D(s)}$), namely, $\widetilde{Q}_{0,L}(s)=\widetilde{Q}(s)$ with
\begin{eqnarray}
\fl \widetilde{Q}(s)&\equiv \frac{s+2\alpha}{s+\kappa+2\gamma}\frac{v^{-1}}{[\sqrt{s/D(s)} +\widehat{\Gamma}(s)] \e^{\sqrt{s/D(s)}L/2}-[\sqrt{s/D(s)} -\widehat{\Gamma}(s)]\e^{-\sqrt{s/D(s)}L/2}}\nonumber \\
\fl & =\frac{1}{2}\frac{  s+2\alpha }{v(s+\kappa+2\gamma)\sqrt{s/D(s)}\calF_S(s)  +(s+\kappa)(s+2\alpha)\calF_C(s) },
\label{bbb}
\end{eqnarray}
and
\begin{equation}
\calF_S(s) =\sinh(\sqrt{s/D(s)}L/2),\quad \calF_C(s) =\cosh(\sqrt{s/D(s)}L/2).
\end{equation}

\noindent (ii) Equation (\ref{bbb}) simplifies even further when $\kappa=0$ (no absorption) and $\alpha=\gamma$, that is, the rate of switching between the velocity states is the same at the ends as in the bulk:
\begin{eqnarray}
\fl \widetilde{Q}(s) =\frac{1}{2}\frac{1}{s\cosh(\sqrt{s/D(s)}L/2)+v\sqrt{s/D(s)} \sinh(\sqrt{s/D(s)}L/2)}.
\label{LTQ}
\end{eqnarray}
Using the property $Q_{\infty}\equiv\lim_{t\rightarrow \infty} Q(t)=\lim_{s\rightarrow 0} s\widetilde{Q}(s)$, we obtain the result
\[Q_{\infty} =\frac{1}{2}\left (1+\frac{\alpha L}{v}\right )^{-1}.\]
In the absence of absorption, we have the conservation equation (\ref{norm}) so that
$Lp_{\infty}+2Q_{\infty}=1$ and hence $p_{\infty}=2\alpha  
Q_{\infty}/v$. Also note that the steady state is independent of the initial position $x_0$.
\medskip

\noindent (iii) Even for finite non-zero values of $\kappa$, the boundaries become totally reflecting in the limit $\gamma\rightarrow \infty$, whereby $j(0,t)=0=j(L,t)$, with $Q_{0,L}(t)\equiv 0$ for all $t$. This is due to the fact that as soon as the RTP hits a walk and enters a bound state, it immediately exits the state without any chance to be absorbed. Equation (\ref{psticky}) then reduces to the solution of the resulting Neumann BVP.

\subsection{First passage time problem}

In the presence of absorption, the conservation equation (\ref{norm}) no longer holds. The associated survival probability is
\begin{equation}
\label{SP}
S(x_0,t)=\int_0^{L}p(x,t)dx+Q_0(t)+Q_L(t).
\end{equation}
Differentiating both sides with respect to $t$ using equations (\ref{JDLa}), (\ref{JDLb}) and the boundary conditions (\ref{stick0}) and (\ref{stick1}), we have
\begin{eqnarray}
\frac{\partial S}{\partial t}&=-\int_0^L\frac{\partial j(x,t)}{\partial x}dx-j(0,t)-\kappa Q_0(t)+j(L,t)-\kappa Q_L(t)\nonumber \\
&=-\kappa[Q_0(t)+Q_L(t)].
\end{eqnarray}
Let $T(x_0)$ denote the FPT for absorption at either end. The corresponding FPT density is
\begin{equation}
f(x_0,t)=-\frac{\partial S}{\partial t}=\kappa[Q_0(t)+Q_L(t)],
\end{equation}
and the corresponding MFPT is
\begin{eqnarray}
\fl \tau(x_0):= \E[ T(x_0)] &= \int_0^{\infty}t f(x_0,t)dt= -\int_0^{\infty}t\frac{\partial S(x_0,t)}{\partial t}dt =\int_0^{\infty} S(x_0,t)dt,
\end{eqnarray}
after integration by parts. Hence,
\begin{eqnarray}
 \tau(x_0)=\widetilde{S}(x_0,0)=-\kappa\lim_{s\rightarrow 0} \frac{\partial}{\partial s}[\widetilde{Q}_0(s)+\widetilde{Q}_L(s)].
 \end{eqnarray}

For simplicity, suppose that $x_0=L/2$. After some algebra, we find that
  \begin{equation}
 \tau(L/2)=\frac{\kappa+2\gamma}{\kappa}\frac{L}{2v}+\frac{\alpha L^2}{4v^2}+\frac{1}{\kappa}.
 \label{tauL}
 \end{equation}
This recovers the result derived in Ref. \cite{Angelani17} when $\alpha=\gamma$. In the limit $\kappa\rightarrow 0$, the MFPT $\tau\rightarrow \infty$ since the sticky boundaries become non-absorbing. On the other hand, in the limit $\kappa\rightarrow \infty$, the RTP is absorbed as soon as it hits one of the boundaries and 
\begin{equation}
\tau(L/2) \rightarrow \tau_{\infty}(L/2)\equiv L/2v+\alpha L^2/4v^2. 
\end{equation}
That is, we recover the MFPT $\tau_{\infty}$ for an RTP starting in the center of the domain with totally absorbing boundaries at either end. A modified version of this result occurs when $\kappa$ is finite and $\gamma=0$, that is, $\tau =\tau_{\infty}+1/\kappa$. Recall that the boundaries are now totally absorbing but the RTP remains in a bound state until it is killed over a mean time interval $1/\kappa$.
Finally, applying the double limit (\ref{kay0}) to equation (\ref{tauL}) gives
 \begin{equation}
 \tau(L/2)=\left (1+\frac{2v}{k_0}\right )\frac{L}{2v}+\frac{\alpha L^2}{4v^2}.
 \label{tauL2}
 \end{equation}

It is also straightforward to determine the splitting probabilities $\pi_{o}(x_0)$ and $\pi_L(x_0)$ for absorption at the ends $x=0$ and $x=L$, respectively. In particular, $\pi_L(x_0)=1-\pi_0(x_0)$ with
\begin{eqnarray}
\pi_0(x_0)&=\widetilde{j}(0,0) =\kappa \widetilde{Q}_0(0)=\frac{(L-x_0)+1/\Gamma(0)}{L+2/\Gamma(0)}\nonumber \\
&=\frac{(L-x_0)+v[\kappa+2\gamma]/(2\alpha \kappa)}{L+2v[\kappa+2\gamma]/(2\alpha \kappa)}.
\label{piL}
\end{eqnarray}
As expected, $\pi_0(L/2)=1/2$ and $\pi_0(x_0)$ is a decreasing function of $x_0$. Moreover, $\pi_0(x_0) \rightarrow 1/2$ in the zero switching limit $\alpha\rightarrow 0$ since $\Gamma(0)\rightarrow 0$. This reflects the initial condition that $\sigma(0)=\pm$ with equal probability.
On the other hand, in the non-sticky limit $\gamma,\kappa\rightarrow \infty$ with $k_0$ fixed in equation (\ref{kay0}), we have $\Gamma(0)\rightarrow 2\alpha k_0/(v[2v+k_0])$ and the splitting probability becomes
\begin{equation}
\pi_0(x_0)=\frac{(L-x_0)+{v^2}/{(\alpha k_0)}+v/(2\alpha)}{L+{2v^2}/{(\alpha k_0)}+v/\alpha}.
\end{equation}
 Finally, note that $\lim_{v\rightarrow 0^+}\pi_0(x_0)=(L-x_0)/L$. This should be interpreted as a singular limit in the sense that if $v=0$ then the RTP cannot move so there is no absorption and $\pi_0(x_0)=0=\pi_L(x_0)$.
   
   \section{Encounter-based model and the occupation time propagator}
   
As we highlighted in the introduction, the absorption mechanism for an RTP with partially absorbing sticky boundaries differs significantly from the non-sticky case, since absorption now occurs from the bound state $B_0$ or $B_L$, see Fig. \ref{fig1}. Therefore, consider the occupation time $A(t)$, which is the amount of time that the RTP spends in the state $B_0\cup B_L$ over the time interval $(0,t)$. In this section we do not distinguish between the time spent in $B_0$ and the time spent in $B_L$. First, suppose that the sticky boundaries are non-absorbing.
Introduce the joint probability density or occupation time propagator
\begin{eqnarray}
\fl \calP_{\sigma}(x,a,t)dxda =\P[x<X(t)<x+dx,a<A(t)<a+da,\sigma(t)=\sigma\in \{+,-\}],
\end{eqnarray}
with $X(0)=x_0$, $\sigma(t)=\pm$ with probability 1/2, and $A(0)=0$.
Since the occupation time only changes at the boundaries, the evolution equation within the bulk of the domain takes the same form as for $p_{\sigma}(x,t)$. In terms of the transformed propagators
\numparts
\begin{eqnarray}
\calP(x,a,t)&=\calP_+(x,a,t)+\calP_-(x,a,t),\\ \calJ(x,a,t)&=v[\calP_+(x,a,t)-\calP_-(x,a,t)],
\end{eqnarray}
\endnumparts
we have
\numparts
 \begin{eqnarray} 
   \label{JPCK1}
   \frac{\partial \calP}{\partial t}   &=&- \frac{\partial \calJ}{\partial x},\\
 \frac{\partial \calJ}{\partial t}   &=&-v^2 \frac{\partial \calP}{\partial x}-2\alpha \calJ .
   \label{JPCK2}
  \end{eqnarray}
  \endnumparts
   
   The nontrivial step is determining the boundary conditions at $x=0,L$. Based on our previous studies of the occupation time propagator for Brownian particles \cite{Bressloff22,Bressloff22a}, we introduce the joint densities 
   \begin{equation}
   \calQ_{0,L}(a,t)da=\P[a<A(t)<a+da , \sigma(t)=B_{0,L}],
   \label{Qdef}
   \end{equation}
    and take
   \numparts
\begin{equation}
\label{calstick0}
\fl \frac{\partial \calQ_0}{\partial a}+\frac{\partial \calQ_0}{\partial t}=-\delta(a)\calQ_0(0,t)-\calJ(0,a,t), \quad \gamma \calQ_0(a,t)=v\calP_+(0,a,t),
\end{equation}
and
\begin{equation}
\label{calstick1}
\fl \frac{\partial \calQ_L}{\partial a}+\frac{\partial \calQ_L}{\partial t}=-\delta(a) \calQ_L(0,t)+\calJ(L,a,t),\quad \gamma \calQ_L(a,t)=v\calP_-(L,a,t).
\end{equation}
\endnumparts
The sum of time derivatives on the left-hand side of equations (\ref{calstick0}) and (\ref{calstick1}) reflects the fact that whenever the RTP is in a bound state, the occupation time $A(t)$ increases at the same rate as the absolute time $t$. We thus have an age-structured model. This type of model typically arises within the context of birth-death processes in ecology and cell biology,
 where the birth and death rates depend on the age of the underlying populations 
\cite{McKendrick25,Foerster59,Iannelli17}. The extra time variable is then the age of an organism or cell, rather than the occupation time. The term involving the Dirac delta function $\delta(a)$ on the right-hand of equations (\ref{calstick0}) and (\ref{calstick1}) ensures that the probability of being stuck at a boundary is zero if the occupation time is zero.
   
Introduce the double Laplace transforms with respect to $t$ and $a$ by setting
\numparts
\begin{eqnarray}
\PP(x,z,s)&=\int_0^{\infty} \e^{-za}\left [\int_0^{\infty} \e^{-st}\calP(x,a,t)dt\right ]da,\\ \QQ(z,s )&=\int_0^{\infty} \e^{-za}\left [\int_0^{\infty} \e^{-st}\calQ(a,t)dt\right ]da.
\end{eqnarray}
\endnumparts
This yields the system of equations
\numparts
\begin{eqnarray}
\label{PPLa}
\fl & D(s)\frac{\partial^2 \PP}{\partial x^2}-s \PP=- \delta(x-x_0) ,\\
\fl &v\PP(0,z,s)=-\frac{2\gamma+s+z}{s+z}\widetilde{\calJ}(0,z,s),\quad v\PP(L,s,z)=\frac{2\gamma+s+z}{s+z}\widetilde{\calJ}(L,s,z),
\label{PPLb}\\
\fl &\gamma \QQ_0(z,s)=\frac{v\PP(0,z,s)+\widetilde{\calJ}(0,z,s)}{2},\quad \gamma \QQ_L(z,s)=\frac{v\PP(L,z,s)-\widetilde{\calJ}(L,z,s)}{2}.
\end{eqnarray}
\endnumparts
 Comparison with the BVP for a constant rate of absorption $\kappa$, see equations (\ref{stickyL}) and (\ref{stickyL2}), shows that
\begin{equation}
\label{loco}
\p(x,s)= \PP(x,z=\kappa,s),\quad \j(x,s)= \widetilde{\calJ}(x,z=\kappa,s) ,
\end{equation}
where $\p(x,s)$ is the solution (\ref{psticky}) etc.

As with the encounter-based formulation of diffusion \cite{Grebenkov19b,Grebenkov20,Grebenkov22,Bressloff22}, the relations (\ref{loco}) have a natural probabilistic interpretation. First, we introduce the stopping time condition
\begin{equation}
\label{TA}
{\mathcal T}=\inf\{t>0:\ A(t) >\widehat{A}\},
\end{equation}
where $\widehat{A}$ is a random variable with probability distribution $\P[\widehat{A}>a]=\Psi(a)$. Note that ${\mathcal T}$ is a random variable that specifies the first absorption time when the RTP is in a bound state, which is identified with the event that $A(t)$ first crosses a randomly generated threshold $\widehat{A}$. The marginal probability density for particle position $X(t) $ is then
\[p^{\Psi}(x,t)dx=\P[X(t) \in (x,x+dx), \ t < {\mathcal T}].\]
Given that $A(t)$ is a nondecreasing process, the condition $t < {\mathcal T}$ is equivalent to the condition $A(t)<\widehat{A}$. This implies that 
\begin{eqnarray}
p^{\psi}(x,t)dx&=\P[X(t) \in (x,x+dx), \ A(t) < \widehat{A}]\nonumber \\
&=\int_0^{\infty} da \ \psi(a)\int_0^{a} da' [\calP(x,a',t)dx],
\end{eqnarray}
where $\psi(a)=-d\Psi(a)/da$. Using the identity
\[\int_0^{\infty}du\ f(u)\int_0^u du' \ g(u')=\int_0^{\infty}du' \ g(u')\int_{u'}^{\infty} du \ f(u)\]
for arbitrary integrable functions $f,g$, it follows that
\begin{eqnarray}
\label{peep}
  p^{\Psi}(x,t)&=\int_0^{\infty}\Psi(a) \calP(x,a,t)da   .
\end{eqnarray}
Using an identical argument, we also have
\begin{eqnarray}
\label{peep2}
 Q^{\psi}_{0,L}(t)&=\int_0^{\infty}\Psi(a) \calQ_{0,L}(a,t)da  .
\end{eqnarray}

In conclusion, the probability densities $p_{\pm}(x,t)$ and bound state probabilities $Q_{0,L}(t)$ for a constant rate of absorption $\kappa$ can be expressed in terms of the Laplace transform of the occupation time propagators $\calP_{\pm}(x,a,t)$ and $\calQ(a,t)$ with respect to the occupation time $a$, since the partially absorbing sticky boundary conditions map to an exponential law for the threshold occupation time $\widehat{A}$. 
The advantage of the probabilistic formulation of a partially absorbing boundary is that one can consider a more general probability distribution $\Psi(a) $ for the occupation time threshold such that 
 \numparts
   \label{coo}
  \begin{eqnarray}
  \p^{\Psi} (x,s)&=\int_0^{\infty} \Psi(a){\mathcal L}_a^{-1}[\PP(x,z,s)]da , \\
  \widetilde{Q}^{\Psi} _{0,L}(s)&=\int_0^{\infty} \Psi(a){\mathcal L}_a^{-1}[\QQ_{0,L}(z,s)]da   .
  \end{eqnarray}
  \endnumparts
  where ${\mathcal L}_a^{-1}$ denotes the inverse Laplace transform
   Hence, we can incorporate a more general model of absorption at sticky boundaries by solving the BVP given by equations (\ref{stickyL}) and (\ref{stickyL2}), reinterpreting the constant absorption rate $\kappa$ as the Laplace variable $z$ that is conjugate to the occupation time, inverting the Laplace transform with respect to $z$, and then evaluating the integrals with respect to $a$, see Fig. \ref{fig1}(b).  

\paragraph{Limit $\gamma \rightarrow \infty$.} It is instructive to understand how the continuous occupation time formalism for sticky boundaries is related to the discrete local time formalism for non-sticky boundaries developed in our previous paper \cite{Bressloff22rtp}. In order to establish such a connection, let $\ell(t)$ denote the number of times that the RTP has hit either end in the time interval $[0,t]$. Assuming that the particle is not in a bound state at time $t$, we can represent the stochastic occupation time in the form
\begin{equation}
A(t)=\sum_{n=1}^{\infty}\tau_n \Theta(\ell(t)-n),
\end{equation}
where $\tau_n$ is the time spent in a bound state following the $n$-th collision with a boundary.  The times $\tau_n$ are identical, identically distributed random variables generated from the exponential density $\gamma\e^{-\gamma \tau}$. In particular, $\E[\tau_n]=\gamma^{-1}$ and $\mbox{Var}[\tau_n]=\gamma^{-2}$. Hence, in the large-$\gamma$ limit, the RTP approximately spends an infinitesimal time $\gamma^{-1}$ in a bound state following each collision, implying that
\begin{equation}
A(t)\rightarrow \gamma^{-1}\ell(t) .
\end{equation}
Under such an approximation, we can set $da=\gamma^{-1}$ in equation (\ref{Qdef}) to give
 \begin{equation}
 \calQ_{0,L}(a,t)da \approx \gamma^{-1} \calQ_{0,L}(\gamma^{-1} \ell,t)= \P[\ell<N(t)<\ell+1 ].
   \label{Qdef2}
   \end{equation}
   Moreover,
   \numparts
   \begin{eqnarray}
   \fl \frac{\partial Q_{0}}{\partial a}\approx \gamma[\calQ_{0}(a+\gamma^{-1},t)-\calQ_{0}(a,t)]=
   v[\calP_+(0,a+\gamma^{-1},t)-\calP_+(0,a,t)],\\
   \fl \frac{\partial Q_{L}}{\partial a}\approx \gamma[\calQ_{L}(a+\gamma^{-1},t)-\calQ_{L}(a,t)]=
   v[\calP_-(L,a+\gamma^{-1},t)-\calP_-(L,a,t)].
   \end{eqnarray}
   \endnumparts
Substituting into equations (\ref{calstick0}) and (\ref{calstick1}), and defining discrete local propagators via 
\begin{equation}
\gamma^{-1}\calP(x,\gamma^{-1}\ell,t)=P(x,\ell,t),\quad \gamma^{-1}\calJ(x,\gamma^{-1}\ell,t)=J(x,\ell,t),
\end{equation}
we obtain the boundary conditions
\numparts
\begin{eqnarray}
J(0,\ell,t) 
&=-v[P_+(0,\ell+1,t )-P_+(L,\ell,t)],\\
 J(L,\ell,t) 
&=v[P_-(L,\ell+1,t)-P_-(L,\ell,t)],
\end{eqnarray}
\endnumparts
with $P_-(L,0,t)=0=P_0(0,0,t)$. Note that $\partial \calQ_0/\partial t=\gamma^{-1}v\partial \calP_+/\partial t$ and $\partial \calQ_L/\partial t=\gamma^{-1}v\partial \calP_-/\partial t$ so that these terms can be dropped in the limit $\gamma \rightarrow \infty$. Following Ref. \cite{Bressloff22rtp}, let us introduce the discrete Laplace transforms 
\begin{equation}
\label{0LT}
\fl  \widetilde{P}(x,w,t|x_0)=\sum_{\ell=0}^{\infty} w^{\ell} P(x,\ell,t|x_0),\quad  \widetilde{J}(x,w,t|x_0)=\sum_{\ell=0}^{\infty} w^{\ell}  J(x,\ell,t|x_0),
\end{equation}
with $w\in [0,1]$, and similarly for $\widetilde{P}_{\pm}(x,w,t)$. The transformed propagator satisfies the boundary conditions
   \begin{equation}
   \label{BCLLT}
\fl \widetilde{J}(0,w,t|x_0)=-\frac{v[1-w]}{w}\widetilde{P}_+(0,w,t|x_0),\ \widetilde{J}(L,w,t|x_0)=\frac{v[1-w]}{w}\widetilde{P}_-(L,w,t|x_0).
\end{equation}
 \endnumparts
 Comparison with the Laplace transform of equations (\ref{calstick0}) and (\ref{calstick0}) with respect to $a$ (after dropping the time derivatives) leads to the approximate mapping
 \begin{equation}
 \frac{vz}{\gamma} \rightarrow\frac{v(1-w)}{w}.
 \end{equation}
 Setting $z=\kappa$ and taking the double limit $\kappa\rightarrow \infty$ and $\gamma \rightarrow \infty$ implies that $k_0=v(1-w)/w$. This recovers one of the results of Ref. \cite{Bressloff22rtp}.

 \setcounter{equation}{0}
\section{MFPT and splitting probabilities for generalized absorption} 

We will illustrate the encounter-based formulation by returning to the FPT problem considered in section 2.2. First, we define the generalized survival probability
\begin{eqnarray}
\fl S^{\Psi}(x_0,t)&=\int_0^{L}p^{\Psi}(x,t)dx+Q^{\Psi}_0(t)+Q^{\Psi}_L(t)\nonumber \\
\fl &=\int_0^{\infty}\Psi(a) \left [\int_0^{L}P(x,a,t)dx+Q_0(a,t)+Q_L(a,t)\right ]da.
\label{SPgen}
\end{eqnarray}
Differentiating both sides with respect to $t$ using equations (\ref{JPCK1}), (\ref{calstick0}) and (\ref{calstick1}) gives
\begin{eqnarray}
\frac{\partial S^{\Psi}}{\partial t}&=\int_0^{\infty} \Psi(a)\bigg [-\int_0^L\frac{\partial \calJ(x,a,t)}{\partial x}dx-\calJ(0,a,t)-\frac{\partial Q_0(a,t)H(a)}{\partial a}\nonumber \\
&\hspace{2cm} +\calJ(L,a,t)-\frac{\partial Q_L(a,t)H(a)}{\partial a}\bigg ]\nonumber \\
 &=-\int_0^{\infty}\psi(a) [Q_0(a,t)+Q_L(a,t)]da.
\end{eqnarray}
Let $T^{\Psi}(x_0)$ denote the FPT for absorption at either end in the case of the general distribution $\Psi(a)$. The corresponding FPT density is
\begin{equation}
f^{\Psi}(x_0,t)=-\frac{\partial S^{\Psi}}{\partial t}=\int_0^{\infty}\psi(a)[Q_0(a,t)+Q_L(a,t)]da,
\end{equation}
and the generalized MFPT is
\begin{eqnarray}
 \tau^{\Psi}(x_0)=-\lim_{s\rightarrow 0} \int_0^{\infty}\psi(a)\left [{\mathcal L}_a^{-1}\frac{\partial}{\partial s}[\QQ_0(z,s)+\QQ_L(z,s)]\right ]da.
 \label{tada}
 \end{eqnarray}
 Similarly, the generalized splitting probabilities are $\pi_L^{\Psi}(x_0)=1-\pi_0^{\Psi}(x_0)$ with
 \begin{eqnarray}
\pi_0^{\Psi}(x_0)&=\lim_{s\rightarrow 0}\int_0^{\infty}\psi(a)  \left [{\mathcal L}_a^{-1}\widetilde{Q}_0(z,s)\right ]da.
\end{eqnarray}

\subsection{MFPT}

 In the previous section we established that the double Laplace transforms $\QQ_{0,L}(z,s)$ are identical to the solutions (\ref{LTQ0}) and (\ref{LTQL}), after identifying the absorption rate $\kappa$ as the Laplace variable $z$. For simplicity, we will focus on the case $x_0=L/2$ so that, from equation (\ref{bbb}) we have $\QQ_{0,L}(z,s)=\QQ(z,s)$ with
 \begin{eqnarray}
\QQ(z,s)&=\frac{1}{2}\frac{  s+2\alpha }{v(s+z+2\gamma)\sqrt{s/D(s)}\calF_S(s)  +(s+z)(s+2\alpha)\calF_C(s)}\nonumber \\
 &=\frac{1}{2}\frac{s+2\alpha}{C_0(s)+zC_1(s)},
\end{eqnarray}
and
\numparts
\begin{equation}
C_0(s)=v(s+2\gamma)\sqrt{s/D(s)}\calF_S(s)+s(s+2\alpha)\calF_C(s),
\end{equation}
\begin{equation}
C_1(s)=v\sqrt{s/D(s)}\calF_S(s)+(s+2\alpha)\calF_C(s).
\end{equation}
\endnumparts
Since $\QQ(z,s)$ has a simple pole with respect to $z$, it is straightforward to find the inverse Laplace transform:
\begin{equation}
\hcalQ(a,s)=\frac{s+2\alpha}{2C_1(s)}\e^{-C_0(s)a/C_1(s)}.
\end{equation}
Substituting into (\ref{tada}) yields the result
\begin{eqnarray}
\fl  \tau^{\Psi}(L/2)&=-2\lim_{s\rightarrow 0} \int_0^{\infty}\psi(a)\frac{\partial}{\partial s}\hcalQ(a,s)=-\lim_{s\rightarrow 0} \frac{\partial}{\partial s}\left [\frac{s+2\alpha}{C_1(s)}\widetilde{\psi}(C_0(s)/C_1(s))\right ].
 \end{eqnarray}
Evaluating the various $s$-derivatives, we obtain the explicit solution
\begin{eqnarray}
\fl \tau^{\Psi}(L/2)&=\left [\frac{ vL}{4\alpha D_0}+\frac{L^2}{8D_0}\right ]\widetilde{\psi}(0)-\left (1+\frac{\gamma vL}{2\alpha D_0}\right )\widetilde{\psi}'(0).
 \end{eqnarray}
 Finally, using the identities $\widetilde{\psi}(0)=\int_0^{\infty}\psi(a)da=1$ and
 $\widetilde{\psi}'(0)=-\int_0^{\infty}a\psi(a)da=\E[a]$ (assuming the first moment is finite),  
 \begin{eqnarray}
\tau^{\Psi}(L/2)&=\frac{ L}{2v}+\frac{\alpha L^2}{4v^2}+\left (1+\frac{\gamma L}{v}\right )\E[a].
 \end{eqnarray}
Thus reduces to equation (\ref{tauL}) for an exponential distribution $\psi(\ell)=\kappa\e^{-\kappa a}$ since $\E[a]=1/\kappa$.

\begin{figure}[t!]
  \raggedleft
   \includegraphics[width=10cm]{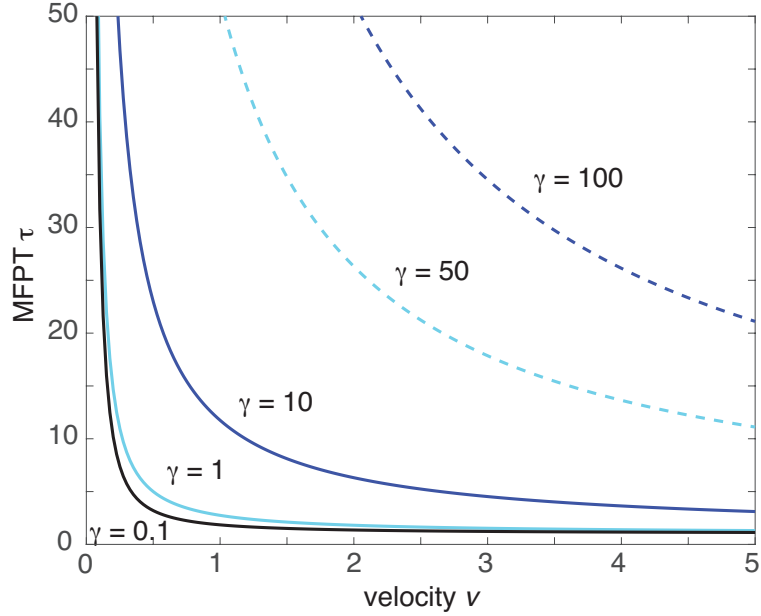}
  \caption{RTP in an interval with partially absorbing sticky boundaries at $x=0,L$. Plot of the MFPT $\tau(L/2)$ as a function of the speed $v$ for a fixed mean $\E[a]=1$ and various values of $\gamma$. The units of space and time are fixed by setting $L=1$ and $\alpha=1$, respectively.}
  \label{fig3}
  \end{figure}

Consistent with our findings for other stochastic processes with generalized absorption \cite{Bressloff22,Bressloff22a,Bressloff22rtp}, the MFPT is only finite for occupation time threshold densities $\psi(a)$ that have a finite first moment. One well known example is the gamma distribution:
\begin{equation}
\psi(a)=\frac{\kappa(\kappa a)^{\mu-1}\e^{-\kappa a}}{\Gamma(\mu)}, \quad \widetilde{\psi}(q)=\left (\frac{\kappa}{\kappa+q}\right )^{\mu},\quad  \mu >0,
\label{gam}
\end{equation}
where $\Gamma(\mu)$ is the gamma function
\begin{equation}
\Gamma(\mu)=\int_0^{\infty}\e^{-t}t^{\mu-1}dt ,\ \mu >0.
\end{equation}
The mean is $\E[a]\equiv -\widetilde{\psi}'(0)=\mu/\kappa$. If $\mu=1$ then we recover the exponential distribution with constant reactivity $\kappa$, whereas the absorption process is non-Markovian for $\mu \neq 1$, since the effective reactivity is $a$-dependent. That is, we can set
\begin{equation}
\label{kaell}
\Psi(a)=\exp\left (-\int_0^{a}\kappa(a')da'\right ),
\end{equation}
where
\begin{equation}
\kappa(a)=\kappa \frac{ (\kappa a)^{\mu-1}\e^{-\kappa a}}{\Gamma(\mu,\kappa a)},
\end{equation} 
and $\Gamma(\mu,z)$ is the upper incomplete gamma function:
\begin{equation}
 \Gamma(\mu,z)=\int_z^{\infty}\e^{-t}t^{\mu-1}dt,\ \mu >0.
\end{equation}
In Fig. \ref{fig3} we show example plots of the MFPT $\tau(L/2)$ as a function of the speed $v$ for different values of the stickiness parameter $\gamma$ and a fixed mean $\E[a]$. It can be seen that the MFPT is an increasing function of $\gamma$, since the RTP spends less fraction of the time in a bound state. On the other hand, $\tau$ converges to the MFPT $\tau_{\infty}$ for totally absorbing boundaries as $\gamma \rightarrow 0$.

 \begin{figure}[t!]
  \raggedleft
   \includegraphics[width=10cm]{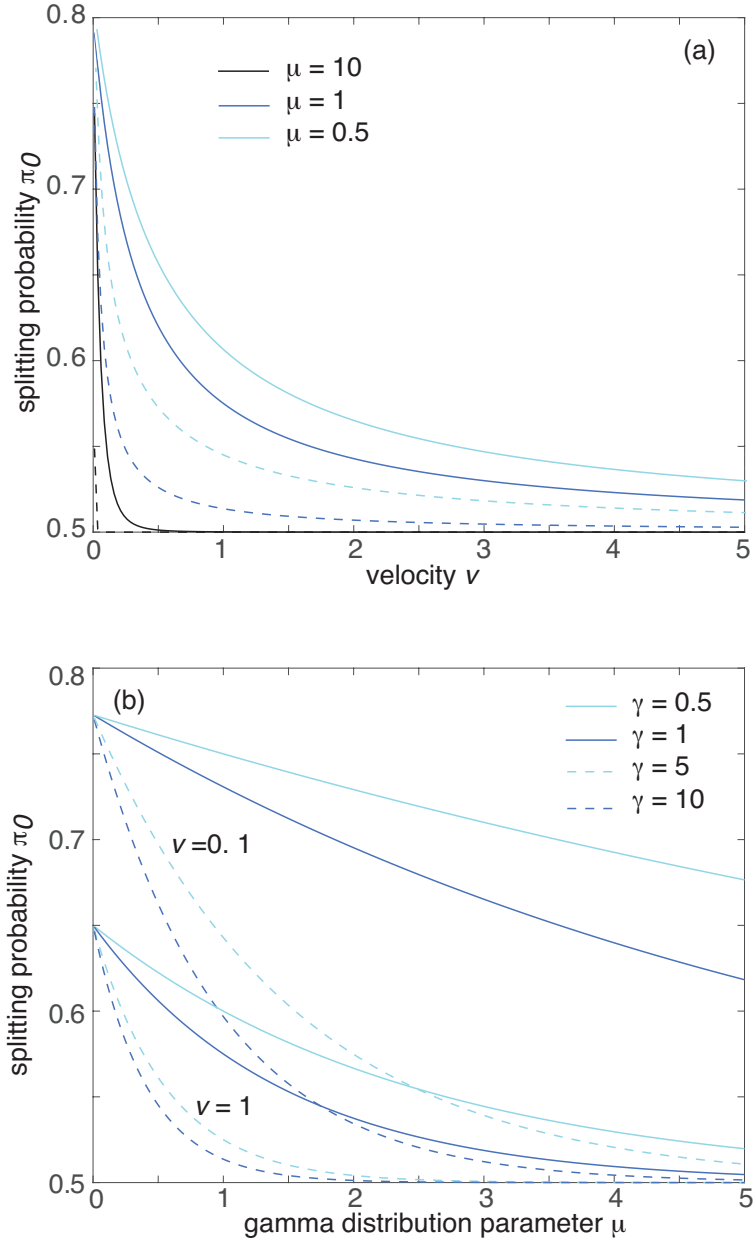}
  \caption{RTP in an interval with partially absorbing, sticky boundaries at $x=0,L$ and $\psi(a)$ given by the gamma distribution (\ref{gam}) with parameters $(\kappa,\mu)$. (a) Plot of the generalized splitting probability $\pi_0^{\Psi}(x_0)$ as a function of the speed $v$ and various values of $\mu$ for $\gamma=1$ (solid curves) and $\gamma=10$ (dashed curves). (b) Corresponding plots of $\pi_0^{\Psi}(x_0)$ as a function of $\mu$ for various $\gamma$ and fixed speeds $v=0.1,1$. The units of space and time are fixed by setting $L=1$ and $\alpha=1$, respectively. We also set $\kappa=1$ and $x_0=0.2$.}
  \label{fig4}
  \end{figure}

\subsection{Splitting probabilities} Turning to the generalized splitting probability $\pi_0^{\Psi}(x_0)$ we have
\begin{eqnarray}
\pi_0^{\Psi}(x_0)&=\int_0^{\infty}\psi(a)  \hcalQ_0(a,0)da=\int_0^{\infty}\psi(a){\mathcal L}^{-1}_a [\widetilde{Q}_0(z,0)]da\nonumber \\
&=\int_0^{\infty}\psi(a){\mathcal L}^{-1}_a \left [\frac{(L-x_0)+\frac{\dis v(z+2\gamma)}{\dis 2\alpha z}}{zL+\frac{\dis v(z+2\gamma)}{\dis \alpha }}\right ]da,
\nonumber \\
&=\frac{1}{\alpha L+v}\int_0^{\infty}\psi(a){\mathcal L}^{-1}_a\left [\frac{c_0+c_1(x_0)z}{z(z+c_2)}\right ]da,
\end{eqnarray}
where
\begin{equation}
\label{cs}
c_0= \gamma v ,\quad c_1(x_0)=\alpha (L-x_0) +v/2,\quad  c_2=\frac{2\gamma v}{\alpha L+v}
\end{equation}
Inverting the Laplace transform shows that
\begin{eqnarray}
\pi_0^{\Psi}(x_0)
&=\frac{1}{\alpha L+v}\int_0^{\infty}\psi(a)\left (\frac{c_0}{c_2} -\frac{c_0-c_1(x_0)c_2}{c_2}\e^{-c_2a}\right )da\nonumber \\
&=\frac{c_0 -(c_0-c_1(x_0)c_2)\widetilde{\psi}(c_2)}{c_2(\alpha L+v)}.
\label{piL2}
\end{eqnarray}
It can be checked that $c_0=c_1(L/2)c_2$ and hence $\pi_0^{\Psi}(L/2)=1/2$. Moreover, in the case of an exponential density $\psi(a)=\kappa\e^{-\kappa a}$ we have $\widetilde{\psi}(z)=\kappa/(\kappa+z)$ and equation (\ref{piL2}) reduces to the solution in (\ref{piL}).

\begin{figure}[t!]
  \raggedleft
   \includegraphics[width=10cm]{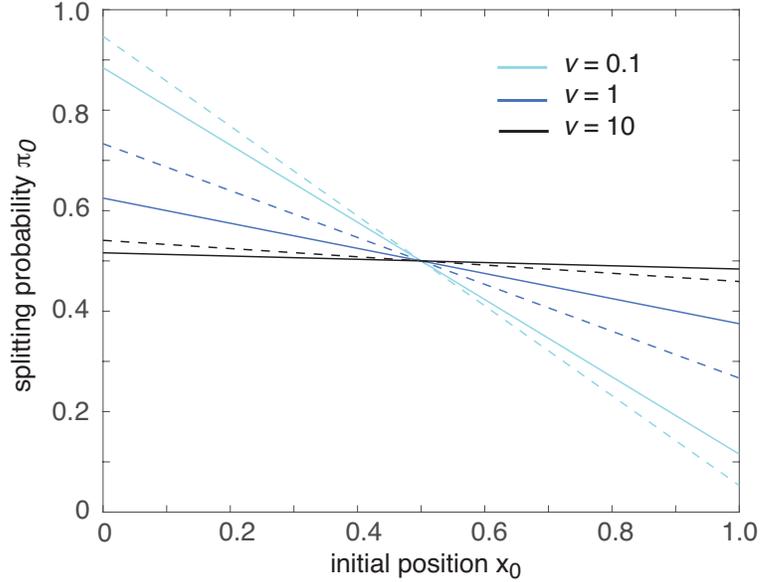}
  \caption{RTP in an interval with partially absorbing sticky boundaries at $x=0,L$. Plot of the MFPT $\tau(L/2)$ as a function of the speed $v$ for a fixed mean $\E[a]=1$ and various values of $\gamma$. The units of space and time are fixed by setting $L=1$ and $\alpha=1$, respectively.}
  \label{fig5}
  \end{figure}

In Fig. \ref{fig4}(a) we plot $\pi_0(x_0)$, $x_0<0.5$, as a function of the velocity $v$ for the gamma distribution (\ref{gam}) for different values of $\mu$ and $\gamma$. It can be seen that $\pi_0(x_0)$ is a monotonically decreasing function of $v$ for fixed $\kappa$, with $\pi_0(x_0)\rightarrow (L-x_0)/L$ as $v \rightarrow 0^+$. $\pi_0$ is also a decreasing function of $\gamma$ and $\mu$. The latter is consistent with the observation that $\psi(a)$ decreases more rapidly (slowly) as a function of the occupation time $a$ when $a <1$ ($a>1$). This is further illustrated in Fig. \ref{fig4}(b). Finally, in Fig. \ref{fig5} we show plots of $\pi_0(x_0)$ as a function of the initial position $x_0$ for various values of $v$ and $\mu$. This shows that when $x_0 >0.5$, the splitting probability $\pi_0(x_0)$ becomes an increasing function of $v$.

\begin{figure}[t!]
  \raggedleft
   \includegraphics[width=13cm]{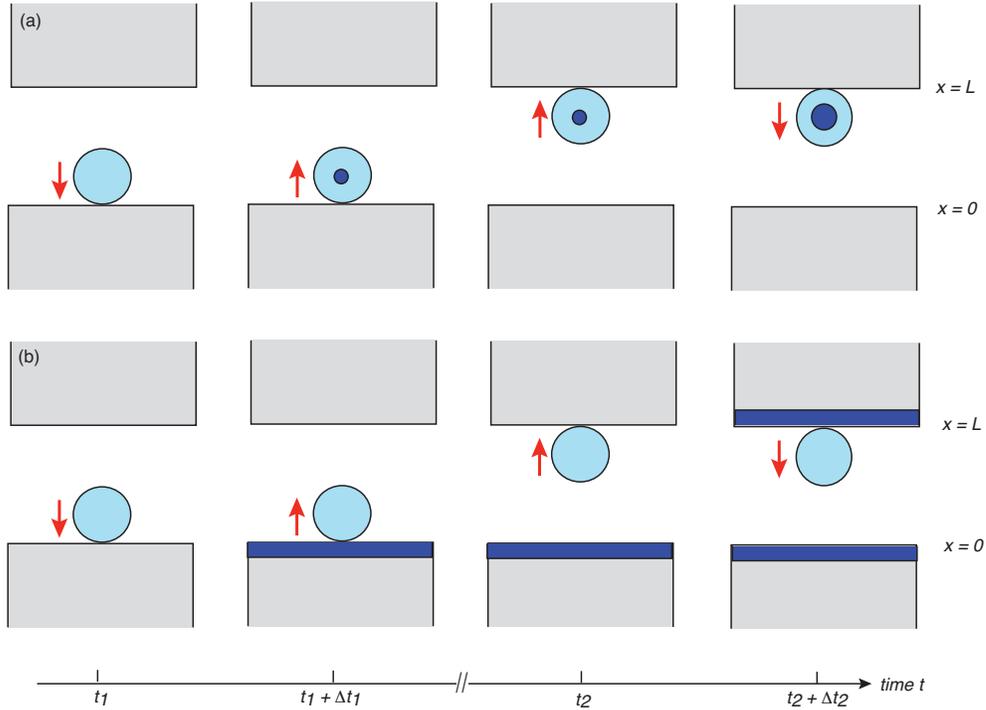}
  \caption{Schematic diagram indicating two different absorption schemes. For the sake of illustration, we consider a sequence of events where the particle first hits the end at $x=0$ and then hits the end at $x=L$.(a) The particle has an internal state $S(t)$ that increases strictly monotonically with the amount of time $A(t)$ spent in contact with either wall. Absorption occurs when the internal state, and hence $A(t)$, crosses a threshold. (b) Each wall has its own internal state denoted by $S_0(t)$ and $S_L(t)$,which is a strictly monotonic function of the amount of time the wall is in contact with the particle, which is specified by the pair of occupation times $A_0(t)$ and $A_L(t)$, respectively. Absorption occurs as soon as one of the internal states crosses its corresponding threshold. In both cases (a) and (b), the value of the internal state is represented by the size of the darker shaded regions.}
  \label{fig6}
  \end{figure}

\section{Pair of independent boundary occupation times}
So far we have assumed that the probability of absorption when in either bound state at time $t$ depends on the total amount of time $A(t)$ the RTP has accumulated in {\em both} states $B_0$ and $B_L$.
A more complicated scenario arises when the probability of absorption at $x=0$ only depends on the occupation time $A_0(t)$ of the bound state $B_0$, and  the probability of absorption at $x=L$ only depends on the occupation time $A_L(t)$ of the bound state $B_L$. One possible physical interpretation of the two different cases is shown schematically in Fig. \ref{fig6}. In Fig. \ref{fig6}(a) the particle has an internal state that is a strictly monotonically increasing function of $A(t)$, whereas in Fig. \ref{fig6}(b) each of the boundaries has its own internal state, $S_0(t)$ and $S_L(t)$, with $S_0(t)$ and $S_L(t)$ a strictly monotonically increasing function of $A_0(t)$ and $A_L(t)$, respectively. (An analogous distinction was highlighted in an encounter-based model of Brownian motion with stochastic resetting \cite{Bressloff22b}.) Which of the two scenarios is more appropriate will depend on the particular application (see the caption of Fig. \ref{fig2}). In the case of a bacterium, scenario (a) is more likely, particularly if the walls are treated with some toxin so that the likelihood of the bacterium being killed increases with its exposure. On the other hand, the nucleation site of a polymer filament may degrade with multiple rounds of nucleation, which is more consistent with scenario (b).

Introducing the vector ${\bf A}(t)=(A_0(t),A_L(t))$, the occupation time propagator becomes
\begin{eqnarray}
\fl \calP_{\sigma}(x,\a,t)dxd\a =\P[x<X(t)<x+dx,\a<{\bf A}(t)<\a+d\a,\sigma(t)=\sigma\in \{+,-\}],
\end{eqnarray}
with $X(0)=x_0$, $\sigma(t)=\pm$ with probability 1/2, and $\A(0)=0$. The evolution equations in the bulk domain are identical in form to equations (\ref{JPCK1}) and (\ref{JPCK2}) for $\calP=\calP_++\calP_-$ and $\calJ=v[\calP_+-\calP_-]$, whereas the boundary conditions at $x=0,L$ become
 \numparts
\begin{equation}
\label{pairQ0}
\fl \frac{\partial \calQ_0}{\partial a_0}+\frac{\partial \calQ_0}{\partial t}=-\delta(a_0)\calQ_0(0,a_L,,t)-\calJ(0,\a,t), \quad \gamma \calQ_0(\a,t)=v\calP_+(0,\a,t),
\end{equation}
with
\begin{equation}
\label{pairQL}
\fl \frac{\partial \calQ_L}{\partial a_L}+\frac{\partial \calQ_L}{\partial t}=-\delta(a_L) \calQ_L(a_0,0,t)+\calJ(L,\a,t),\quad \gamma \calQ_L(\a,t)=v\calP_-(L,\a,t),
\end{equation}
\endnumparts
and
   \begin{equation}
   \calQ_{0,L}(\a,t)d\a=\P[\a<\A(t)<\a+d\a , \sigma(t)=B_{0,L}].
   \label{pairQdef}
   \end{equation}
   Laplace transforming with respect to $t$ and $\a=(a_0,a_L)$ by setting
\numparts
\begin{eqnarray}
\PP(x,\z,s)&=\int_0^{\infty} \e^{-\z\cdot \a}\left [\int_0^{\infty} \e^{-st}\calP(x,\a,t)dt\right ]d\a,\\ \QQ(\z,s )&=\int_0^{\infty} \e^{-\z\cdot \a}\left [\int_0^{\infty} \e^{-st}\calQ(\a,t)dt\right ]d\a,
\end{eqnarray}
\endnumparts
with $\z=(z_0,z_L)$,
we obtain the BVP
\numparts
 \begin{eqnarray}
\label{pairstickyL}
\fl &D(s)\frac{\partial^2 {\PP}}{\partial x^2}-s  {\PP}=- \delta(x-x_0),\\
\label{pairsticky2}
\fl &\left  . \frac{\partial \PP(x,\z,s)}{\partial x}\right |_{x=0}=\Gamma(s,z_0)\PP(0,\z,s),\quad \left  .  \frac{\partial \PP(x,\z,s)}{\partial x}\right |_{x=L}=-\Gamma(s,z_L)\PP(L,\z,s),
\end{eqnarray}
\endnumparts
with
\begin{equation}
\label{Gz}
\Gamma(s,z)=\frac{s+z}{v}\frac{s+2\alpha}{s+z+2\gamma}.
\end{equation}
That is, $\PP(x,\z,s)$ satisfies the BVP for partially absorbing sticky boundaries at $x=0$ and $x=L$ with constant absorption rates $z_0$ and $z_L$, respectively.

The next step is to specify the absorption mechanism. Since we have two occupation times $A_0(t)$ and $A_L(t)$, we introduce
 a corresponding pair of  independent random local time thresholds $\widehat{A}_{0}$ and $\widehat{A}_{0}$ such that
\begin{equation}
 \P[\widehat{A}_{0}>\ell]\equiv \Psi_0(a),\quad \P[\widehat{A}_{L}>\ell]\equiv \Psi_L(a).
 \end{equation}
 Absorption occurs as soon as one of the occupation times  exceeds its corresponding threshold, which occurs at the FPT time 
 \begin{equation}
 \calT=\min\{\tau_{0},\tau_L\},\quad \tau_{0,L}=\inf\{t>0:\ A_{0,L}(t) >\widehat{A}_{0,L}\}.
 \end{equation}
 Since the occupation time thresholds are statistically independent, the relationship between the marginal probability density $p^{\Psi}(x,t)$ and $\calP(x,\a,t)$ can be established as follows:
\begin{eqnarray}
\fl & p^{\Psi}(x,t)dx=\P[X(t) \in (x,x+dx), \ t < {\mathcal T}_j]\nonumber\\
 \fl&=\P[X(t) \in (x,x+dx), \ A_0(t)< \widehat{A}_{0},\ A_L(t)< \widehat{A}_{L}]\nonumber\\
\fl &=\int_0^{\infty} da\,\psi_{0}(a) \int_0^{\infty} da^{\prime} \psi_L(a^{\prime}) \P[X(t) \in (x,x+dx), \ A_0(t) < a, \ A_L(t) <a^{\prime}]\nonumber\\
 \fl &=\int_0^{\infty} da\, \psi_{0} (a) \int_0^{\infty} da^{\prime} \psi_L(a^{\prime})\int_0^{\ell} d\hat{a}  \int_0^{a^{\prime}}d\hat{a}^{\prime}[\calP(x,\hat{a} ,\hat{a} ',t)dx],
\end{eqnarray}
with $\psi_0(a)=-\partial_{a}\Psi_0(a)$ and $\psi_L(a)=-\partial_{a}\Psi_L(a)$.
Reversing the orders of integration yields the result
\begin{eqnarray}
p^{\Psi}(x,t)&=\int_0^{\infty}da\, \Psi_{0}(a) \int_0^{\infty} da^{\prime}\Psi_L(a^{\prime}) \calP(x,a,a^{\prime},t)\nonumber \\
&=\int_0^{\infty}da\, \Psi_{0}(a) \int_0^{\infty} da^{\prime}\Psi_L(a^{\prime}) {\mathcal L}^{-1}_a{\mathcal L}^{-1}_{a'}[\PP(x,\z,t)],
\label{mbob}
\end{eqnarray}
with $\PP(x,\z,t)$ the solution to the BVP with constant absorption rates $\z=(z_0,z_L)$. Using similar arguments,
\begin{eqnarray}
\label{pairpeep2}
 Q^{\Psi}_{0,L}(t)&=\int_0^{\infty}da\,\Psi(a) \int_0^{\infty}da'\Psi(a') \calQ_{0,L}(\a,t) \nonumber \\
 &=\int_0^{\infty}da\,\Psi(a) \int_0^{\infty}da'\Psi(a') {\mathcal L}^{-1}_a{\mathcal L}^{-1}_{a'}[\QQ_{0,L}(\z,t)].
\end{eqnarray}

\begin{figure}[b!]
  \raggedleft
   \includegraphics[width=10cm]{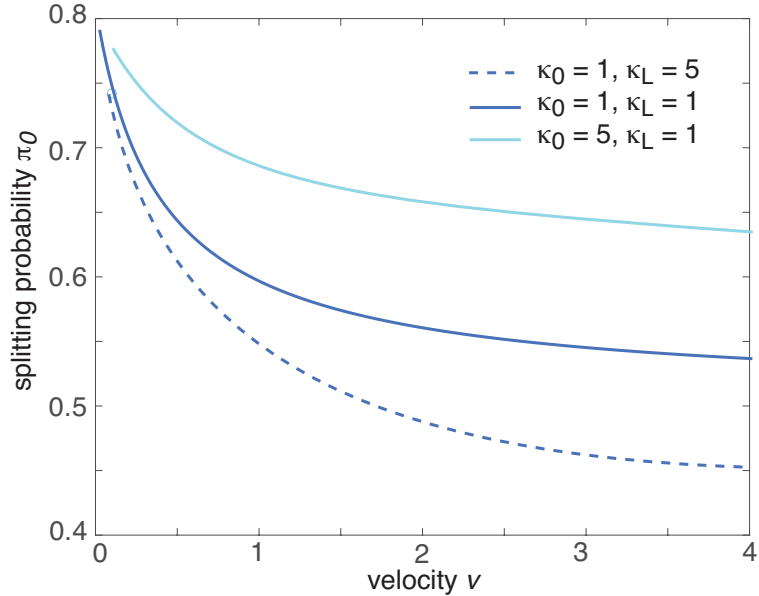}
  \caption{Absorption based on a pair of occupation times $A_0(t)$ and $A_L(t)$ at the walls $x=0$ and $x=L$. The corresponding threshold distributions $\psi_0(a)$ and $\psi_L(a)$ are taken to be gamma distributions with the parameters $(\kappa_0,\mu)$ and $(\kappa_L,\mu)$, respectively. Plot of the generalized splitting probability $\pi_0^{\Psi}(x_0)$ as a function of the speed $v$ for $\mu = 0.5$ and different combinations of $\kappa_0$ and $\kappa_L$. The units of space and time are fixed by setting $L=1$ and $\alpha=1$, respectively. Other parameters are $\gamma=1$ and $x_0=0.2$.}
  \label{fig7}
  \end{figure}

The encounter-based formulation now proceeds along analogous lines to previous sections. First, we solve the BVP of equations (\ref{pairstickyL}) and (\ref{pairsticky2}) to determine $\PP(x,\z,t)$ and $\QQ_{0,L}(\z,t)$. The results are then substituted into equations (\ref{mbob}) and (\ref{pairpeep2}), respectively. The main difficulty arises in performing the double inverse Laplace transform. In order to illustrate this, we consider the generalized splitting probability at $x=0$.
Following section 4, we define the generalized survival probability
\begin{eqnarray}
\label{pairSPgen}
\fl S^{\Psi}(x_0,t)&=\int_0^{L}p^{\Psi}(x,t)dx+Q^{\Psi}_0(t)+Q^{\Psi}_L(t) \\
\fl &=\int_0^{\infty}da\, \Psi_0(a) \int_0^{\infty}da' \Psi_L(a')\left [\int_0^{L}P(x,\a,t)dx+Q_0(\a,t)+Q_L(\a,t)\right ].\nonumber
\end{eqnarray}
Differentiating both sides with respect to $t$ using equations (\ref{JPCK1}), (\ref{pairQ0}) and (\ref{pairQL}) gives
\begin{eqnarray}
\fl \frac{\partial S^{\Psi}}{\partial t}&=\int_0^{\infty}da\, \Psi_0(a) \int_0^{\infty}da' \Psi_L(a')\bigg [-\int_0^L\frac{\partial \calJ(x,a,t)}{\partial x}dx-\calJ(0,\a,t)\nonumber \\
\fl &\hspace{2cm} -\frac{\partial Q_0(\a,t)H(a_0)}{\partial a_0}+\calJ(L,\a,t)-\frac{\partial Q_L(\a,t)H(a_L)}{\partial a_L}\bigg ] \\
\fl  &=-\int_0^{\infty}da\, \psi_0(a) \int_0^{\infty}da' \Psi_L(a')Q_0(a,t)-\int_0^{\infty}da\, \Psi_0(a) \int_0^{\infty}da' \psi_L(a')Q_L(\a,t).\nonumber
\end{eqnarray}
It immediately follows that
 \begin{eqnarray}
\pi_0^{\Psi}(x_0)&=\lim_{s\rightarrow 0}\int_0^{\infty}da\, \psi_0(a) \int_0^{\infty}da' \Psi_L(a') {\mathcal L}^{-1}_a{\mathcal L}^{-1}_{a'}\left [\widetilde{Q}_0(\z,s)\right ].
\label{pisu}
\end{eqnarray}
Extending the analysis of section 2 to the case of different absorption rates at the two ends, we find that
\begin{eqnarray}
\fl   \QQ_0(\z,s) 
 &= \frac{s+2\alpha}{s+z_0+2\gamma}\frac{1/v}{\sqrt{s/D(s)}+\Gamma(s,z_0)}\frac{e^{\sqrt{s/D(s)}(L-x_0)}+\Lambda(s,z_L)\e^{-\sqrt{s/D(s)}(L-x_0)}}{ \e^{\sqrt{s/D(s)}L}-\Lambda(s,z_0)\Lambda(s,z_L)\e^{-\sqrt{s/D(s)}L}}\nonumber \\
 \fl
\label{pairLTQ0}
\end{eqnarray}
\begin{equation}
\label{pairpl}
 \Lambda(s,z)=\frac{\sqrt{s/D(s)}-\Gamma(s,z)}{\sqrt{s/D(s)}+\Gamma(s,z)}.\end{equation}
Hence,
\begin{eqnarray}
\QQ_0(\z,0) =\frac{1}{v}\frac{2\alpha}{z_0+2\gamma}\frac{1+\Gamma(0,z_L)(L-x_0)}{\Gamma(0,z_0)+\Gamma(0,z_L)+\Gamma(0,z_0)\Gamma(0,z_L)L}.
\end{eqnarray}
Substituting for $\Gamma(0,z)$ using equation (\ref{Gz}) yields the analog of equation (\ref{piL}),
\begin{eqnarray}
\QQ_0(\z,0)=\frac{1}{z_0}\frac{(L-x_0)+v\frac{\dis z_L+2\gamma}{\dis 2\alpha z_L}}{L+\frac{\dis v}{\dis 2\alpha}\left[\frac{\dis z_0+2\gamma}{ \dis z_0}+\frac{\dis z_L+2\gamma}{\dis 2\alpha z_L}\right ]}.
\label{QQsu}
\end{eqnarray}
Plugging equation (\ref{QQsu}) into (\ref{pisu}) gives, after some algebra,
\begin{eqnarray}
\fl \pi_0^{\Psi}(x_0)&=\frac{1}{\alpha L+v}\int_0^{\infty}da\, \psi_0(a) \int_0^{\infty}da' \Psi_L(a') {\mathcal L}^{-1}_a{\mathcal L}^{-1}_{a'}\left [\frac{c_0+c_1(x_0)z_L}{z_0z_L+c_2(z_0+z_L)/2)}\right ]
\label{psiu2}
\end{eqnarray}
with the coefficients $c_0$, $c_1(x_0)$ and $c_2$ defined in (\ref{cs}).

The double inverse Laplace transform is evaluated in the Appendix A, and leads to the result
\begin{eqnarray}
\fl &\pi_0^{\Psi}(x_0) \\
\fl &=\frac{1}{\alpha L+v}\int_0^{\infty}da\, \psi_0(a) \int_0^{\infty}da' [c_0\Psi_L(a') +c_1(x_0)\psi_L(a')] \e^{-c_2(a+a')/2} I_0(c_2\sqrt{a' a}), \nonumber
\label{psiu4}
\end{eqnarray}
where $I_0$ is a modified Bessel function of the first kind. Note that $I_0(x)\sim \e^{x}/\sqrt{2\pi x}$ in the large-$x$ limit so that
\[\e^{-c_2(a+a')/2} I_0(c_2\sqrt{a' a})\sim \e^{-c_2(\sqrt{a}-\sqrt{a'})^2/2},\quad a,a'\rightarrow \infty,\]
and we expect $\pi_0^{\Psi}(x_0)$ to be well-defined, at least for distributions with finite moments. For the sake of illustration, suppose that $\psi_{0}(a)$ and $\psi_{L}(a)$ are gamma distributions with parameters $(\kappa_0,\mu)$ and $(\kappa_L,\mu)$, respectively. Numerically evaluating the double integral in equation (\ref{psiu2}) for $\mu=0.5$, we plot $\pi_0^{\Psi}(x_0)$ as a function of the speed $v$ for different combinations of $\kappa_0,\kappa_L$. As expected, the splitting probability is increased when $\kappa_0>\kappa_L$ and decreased when $\kappa_0<\kappa_L$. As a useful check, we find that the numerically calculated curve for $\kappa_0=\kappa_L$ and $\mu=1$ is in excellent agreement with the corresponding analytical curve shown in Fig. \ref{fig5}. (In appendix A we also derive a series representation for $\pi_0^{\Psi}(x_0)$ that is valid when $\mu <1$.)

\section{Discussion}  

In this paper we continued our work on encounter-based models of stochastic processes with partially absorbing boundaries. Previous studies have focused on (passive) Brownian particles \cite{Grebenkov20,Grebenkov22,Bressloff22,Bressloff22a,Bressloff22b}. However, as we showed here and in Ref. \cite{Bressloff22rtp}, the same approach can be applied to RTPs. The basic framework was illustrated in Fig. \ref{fig1} and can be summarized in a more general form as follows: 
\medskip

\noindent (i) Solve the BVP for the probability density of the given stochastic process in the case of a constant absorption rate at each distinct boundary.
\medskip

\noindent (ii) Reinterpret each absorption rate as a Laplace variable conjugate to a local or occupation time that specifies the amount of contact between the particle and the corresponding boundary.
\medskip

\noindent (iii) Invert the Laplace transform with respect to each Laplace variable and evaluate the resulting weighted integral with respect to the boundary encounter times.
\medskip

\noindent The analytical tractability of the method crucially depends on whether or not the BVP has an explicit solution. If the latter is known, then the inverse Laplace transforms can be evaluated using Laplace tables in simple cases, or a combination of spectral theory and contour integration in more complicated examples. Numerical methods such as fast Fourier transforms can also be used.

We conclude that, from a mathematical perspective, extending our analysis to a higher-dimensional version of an RTP
or to an active Brownian particle (ABP) will require solving the full evolution equations. This is a challenging problem except in certain limiting cases such as steady-state or short times, see Refs. \cite{Mori20,Santra20,Santra20a} for RTPs and Refs. \cite{Lee13,Majumdar20,Wagner17,Basu19,Wagner22} for ABPs. We end by indicating how absorption can be included into the standard model of an ABP confined to a 2D channel $\Omega\subset \R^2$ of width $L$ in the $x$ direction and of infinite extension in the $y$ direction, analogous to one interpretation of Fig. \ref{fig2}. Let $\X(t)\in \Omega$ and $\Theta(t) \in [-\pi,\pi]$ denote the position and orientation of the particle at time $t$. These stochastic variables are taken to evolve according to an overdamped Langevin equation of the form
\begin{equation}
d\X(t)=v_0 \n(\Theta(t))+\sqrt{2\overline{D}}d\overline{\bf W}(t),\quad d\Theta(t) =\sqrt{2D}dW(t),
\end{equation}
where $\X(t)=(X(t),Y(t))$, $\overline{\bf W}(t)=(\overline{W}_x(t),\overline{W}_y(t))$ and $\n(\theta)=(\cos \theta,\sin \theta)$. The stochastic variables $\overline{W}_x(t),\overline{W}_y(t),{W}(t)$ are independent Wiener processes, $v_0$ is the speed of the particle, $\overline{D}$ is the translational diffusivity and $D$ is the rotational diffusivity. For simplicity, we will neglect translational diffusion by setting $\overline{D}=0$. Let $p(x,y,\theta,t)$ denote the probability density for the triplet $(X(t),Y(t),\Theta(t))$. The density evolves according to the Fokker-Planck equation
\begin{equation}
\fl \frac{\partial p(\x,\theta,t)}{\partial t}=-v_0\n(\theta)\cdot \nabla p(\x,\theta,t)+D\frac{\partial^2p(\x,\theta,t)}{\partial \theta^2},\quad \x\in \Omega, \quad \theta \in [-\pi,\pi].
\end{equation}
Given the translation invariance in the $y$ direction, we assume that $p$ is independent of $y$ so that the Fokker-Planck equation reduces to the quasi-one dimensional form:
\begin{equation}
\fl \frac{\partial p(x,\theta,t)}{\partial t}=-v_0\cos\theta \frac{\partial p(x,\theta,t)}{\partial x}+D\frac{\partial^2p(x,\theta,t)}{\partial \theta^2},\quad x\in (0,L),\quad \theta \in [-\pi,\pi].
\label{FP}
\end{equation}

The particle will hit the wall at $x=0$ if it is traveling to the left ($\cos \theta <0$), whereas it will hit the wall at $x=L$ if it is traveling to the right ($\cos \theta >0)$. As soon as it hits the wall its linear velocity drops to zero but its orientation will continue to diffuse. The particle remains stuck at the wall until the orientation crosses one of the vertical directions, after which it reenters the bulk domain, or it is permanently killed at some rate $\kappa_0$. Let $Q_0(\theta,t)$ denote the probability density that the particle is attached to the wall at $x=0$ and has orientation $\theta$ ($\cos \theta <0$). Then
\begin{equation}
\fl \frac{\partial Q_0(\theta, t)}{\partial t}=D\frac{\partial^2Q_0(\theta,t)}{\partial \theta^2} 
-v_0\cos \theta  p(0,\theta,t) -\kappa_0Q_0(\theta, t) ,\ \theta \in {\mathcal I},
\label{q0}
\end{equation}
with ${\mathcal I}\equiv (-\pi,-\pi/2)\cup (\pi/2,\pi)$.
The additional term $-\kappa_0 Q_0(\theta,t)$ on the right-hand side of equation (\ref{q0}) represents killing at a constant rate $\kappa_0$. This is analogous to the corresponding term in equation (\ref{stick0}) for an RTP. Indeed, we can interpret $Q_0(t)=\int_{{\mathcal I}}Q_0(\theta,t)d\theta$ as the probability of being in the bound state $B_0$ at $x=0$. Equation (\ref{q0}) is 
supplemented by the absorbing boundary conditions $Q_0(\pm \pi/2,t)=0$, which signal the reinsertion of the particle into the bulk domain. The absorbing boundary conditions mean that the net flux from the left-hand wall back into the bulk is 
\begin{eqnarray}
\label{J0}
\fl v_0 \cos \theta p(0,\theta,t)&=\calJ_0(\theta,t)\\
&\equiv D\frac{\partial Q_0(\pi/2,t)}{\partial \theta}\delta(\theta-\pi/2-\epsilon)-D\frac{\partial Q_0(-\pi/2,t)}{\partial \theta}\delta(\theta+\pi/2+\epsilon),\nonumber
\end{eqnarray}
where $0<\epsilon \ll 1$. The small parameter $\epsilon$ is introduced to avoid the singularities at $\pm \theta =\pi/2$. However, the resulting solution is well defined in the limit $\epsilon \rightarrow 0$.
Similarly, the probability density $Q_L(\theta,t)$ that the particle is attached to the wall at $x=L$ and has orientation $\theta$ ($\cos \theta >0$) evolves according to the equation
\begin{equation}
\label{qL}
\fl \frac{\partial Q_L(\theta, t)}{\partial t}=D\frac{\partial^2Q_L(\theta,t)}{\partial \theta^2} 
+v_0\cos \theta  p(L,\theta,t) -\kappa_0 q_L(\theta,t),\ \theta \in (-\pi/2,\pi/2),
\end{equation}
with $Q_L(\pm \pi/2,t)=0$ and
\begin{eqnarray}
\label{JL}
\fl v_0 \cos \theta p(L,\theta,t)&=\calJ_L(\theta,t)\\
&=-D\frac{\partial Q_L(\pi/2,t)}{\partial \theta}\delta(\theta-\pi/2-\epsilon)+D\frac{\partial Q_L(-\pi/2,t)}{\partial \theta}\delta(\theta+\pi/2+\epsilon).\nonumber 
\end{eqnarray}

Given a solution of equations (\ref{FP})--(\ref{JL}), possibly after Laplace transforming with respect to $t$, a generalized model of absorption could then be introduced by setting $\kappa_0=\kappa_L=\kappa$ and reinterpreting $\kappa$ as a Laplace variable conjugate to the total time spent attached to the walls. One could also introduce separate occupation times at the two ends.

\setcounter{equation}{0}
\renewcommand{\theequation}{A.\arabic{equation}}
\section*{Appendix A: Evaluation of equation (\ref{psiu2})}

Inverting the Laplace transform with respect to $z_L$ in equation (\ref{psiu2}), we have
\begin{eqnarray}
\fl &\pi_0^{\Psi}(x_0)=\frac{1}{\alpha L+v}\int_0^{\infty}da\, \psi_0(a) \int_0^{\infty}da' \Psi_L(a') \nonumber \\
\fl &\hspace{2cm}\times {\mathcal L}^{-1}_a \left \{\frac{1}{z_0+c_2/2}\left [\left (c_0-c_1(x_0)\phi(z_0)\right )\e^{-\phi(z_0)a'} +c_1(x_0)\delta(a')\right ]\right \},
\label{psiu3}
\end{eqnarray}
with
\begin{equation}
\label{phi}
\phi(z_0)=\frac{c_2 z_0/2}{z_0+c_2/2}.
\end{equation}
Defining
\begin{eqnarray}
F(a,a')\equiv  {\mathcal L}^{-1}_a \left \{\frac{1}{z_0+c_2/2} \left (c_0-c_1(x_0)\phi(z_0)\right )\e^{-\phi(z_0)a'} \right \},
\end{eqnarray}
and noting that
\begin{equation}
\left (c_0-c_1(x_0)\phi(z_0)\right )\e^{-\phi(z_0)a'}=\left [c_0+c_1(x_0)\frac{d}{da'}\right ]\e^{-\phi(z_0) a'},
\end{equation}
we have
\begin{eqnarray}
F(a,a')\equiv  \left [c_0+c_1(x_0)\frac{d}{da'}\right ]{\mathcal L}^{-1}_a \left \{\frac{1}{z_0+c_2/2} \e^{-\phi(z_0)a'} \right \}.
\end{eqnarray}
Using a standard table of Laplace transforms, we note that
\begin{eqnarray}
\fl {\mathcal L}^{-1}_a \left \{\frac{1}{z_0+c_2/2} \e^{-\phi(z_0)a'}\right \}&={\mathcal L}^{-1}_a\left \{\frac{1}{z_0+c_2/2}\exp\left [-\left (\frac{c_2}{2}-\frac{(c_2/2)^2}{(z_0+c_2/2)}\right )a' \right ]\right \}\nonumber \\
&=\e^{-c_2(a+a')/2} I_0(c_2\sqrt{a' a}).
\end{eqnarray}
Inverting the Dirac delta function term in (\ref{psiu3}) yields a contribution of the form $c_1(x_0) \delta(a') \e^{-c_2a/2}$. Hence, equation (\ref{psiu3}) reduces to the double integral
\begin{eqnarray}
\fl &\pi_0^{\Psi}(x_0)\nonumber \\
\fl &=\frac{1}{\alpha L+v}\int_0^{\infty}da\, \psi_0(a) \int_0^{\infty}da' \Psi_L(a') \bigg (\left [c_0+c_1(x_0)\frac{d}{da'}\right ]\e^{-c_2(a+a')/2} I_0(c_2\sqrt{a' a})\nonumber \\
\fl &\hspace{6cm} +c_1(x_0) \delta(a') \e^{-c_2a/2}\bigg )
\label{A}\\
\fl & =\frac{1}{\alpha L+v}\int_0^{\infty}da\, \psi_0(a) \int_0^{\infty}da' [c_0\Psi_L(a') +c_1(x_0)\psi_L(a')] \e^{-c_2(a+a')/2} I_0(c_2\sqrt{a' a}),\nonumber 
\end{eqnarray}
after an integration by parts, and we obtain the result (\ref{psiu4}).

Further analytical progress can be made when $\mu <1$. Substituting the infinite series expansion
\begin{equation}
I_0(c_2\sqrt{a' a})=\sum_{j=0}^{\infty} \frac{1}{j!j!} \left (\frac{c_2}{2}\right )^{2j}(a'a)^j.
\end{equation}
into equation (\ref{A}) and assuming that the order of summation and integration can be reversed, we have
\begin{eqnarray}
\fl \pi_0^{\Psi}(x_0) &=\frac{1}{\alpha L+v}\sum_{j=0}^{\infty} \frac{1}{j!j!}\left (\frac{c_2}{2}\right )^{2j} \int_0^{\infty} a^j\psi_0(a) \e^{-c_2 a/2}da\nonumber \\
\fl &\hspace{4cm}\times \int_0^{\infty}(a')^j[c_0\Psi_L(a') +c_1(x_0)\psi_L(a')]  \e^{-c_2 a'/2} da' \nonumber \\
\fl &=\frac{1}{\alpha L+v}\sum_{j=0}^{\infty} \frac{1}{j!j!}\left (\frac{c_2}{2}\right )^{2j} \widetilde{\psi}_0^{(j)}(c_2/2))   \bigg [c_0\widetilde{\Psi}_L^{(j)}(c_2/2) + c_1(x_0)\widetilde{\psi}_L^{(j)}(c_2/2)\bigg]  ,
\end{eqnarray}
with $\widetilde{\psi}^{(j)}(q)=d^j\widetilde{\psi}(q)/dq^j$ and
$\widetilde{\Psi}(q)=(1-\widetilde{\psi}(q))/q$.
 In the case of the gamma distribution
\begin{eqnarray}
\widetilde{\psi}_0^{(j)}(q)=\left (\frac{\kappa_0}{\kappa_0+q}\right )^{\mu}(-1)^j\mu(\mu+1)\ldots (\mu+j-1)(\kappa_0+q)^{-j}
\end{eqnarray}
and similarly for $\widetilde{\psi}_L^{(j)}(q)$. Hence,
\begin{eqnarray}
\pi_0^{\Psi}(x_0) 
&=\frac{1}{\alpha L+v}\frac{2c_0}{c_2}\left (\frac{\kappa_0}{\kappa_0+q}\right )^{\mu}\sum_{j=0}^{\infty} C(j,\mu)\left (\frac{c_2/2}{\kappa_0+c_2/2}\right )^{j} \nonumber \\
&\quad  -\frac{1}{\alpha L+v}\left (c_1(x_0)-\frac{2c_0}{c_2}\right )\left (\frac{\kappa_0}{\kappa_0+c_2/2}\right )^{\mu}\left (\frac{\kappa_L}{\kappa_L+c_2/2}\right )^{\mu}\nonumber \\
 &\qquad \times\sum_{j=0}^{\infty} C(j,\mu)^2\left (\frac{c_2/2}{\kappa_0+c_2/2}\right )^{j} \left (\frac{c_2/2}{\kappa_L+c_2/2}\right )^{j}  ,
\end{eqnarray}
with
\begin{equation}
C(j,\mu)=\frac{(-1)^j\mu(\mu+1)\ldots (\mu+j-1)}{j!}.
\end{equation}
The series representation is clearly convergent when $0<\mu <1$ as $|C(j,\mu)|<1$.


\section*{References}
 
\begin{thebibliography}{9}

\bibitem{Angelani14} Angelani L, Di Lionardo R and Paoluzzi M 2014 First-passage
time of run-and-tumble particles {\em Eur. Phys. J. E} {\bf 37} 59
 

\bibitem{Angelani15} Angelani L 2015 Run-and-tumble particles, telegrapher's equation
and absorption problems with partially reflecting boundaries
{\em J. Phys. A: Math. Theor.} {\bf 48}, 495003 

\bibitem{Angelani17} Angelani L 2017 Confined run-and-tumble swimmers in one dimension {\em J. Phys. A} {\bf 50} 325601  

\bibitem{Basu19} Basu U, Majumdar S N, Rosso A and Schehr G 2019 Active Brownian motion in two dimensions {\em Phys. Rev. E} {\bf 100} 062116

 \bibitem{Dor19}  Ben Dor Y, Woillez E, Kafri Y, Kardar M and Solon A P 2019
Ramifications of disorder on active particles in one dimension
{\em Phys. Rev. E} {\bf 100} 052610  
  
 \bibitem{Bechinger16} Bechinger C, Di Leonardo R, Lowen H, Reichhardt C, Volpe G and Volpe G 2016 Active particles in complex and crowded environments
{\em Rev. Mod. Phys.} {\bf 88} 045006
  
  \bibitem{Berg04} Berg H C 2004 {\em E. Coli in Motion}, New York, Springer
  
  
  
\bibitem{Bressloff11} Bressloff P C and Newby J M 2011 Quasi-steady state analysis of motor-driven transport on a two-dimensional microtubular network. {\em Phys. Rev. E} {\bf 83} 061139 (2011).

\bibitem{Bressloff13} Bressloff P C and Newby J M 2013 Stochastic models of intracellular transport (Review) {\em Rev. Mod. Phys.} {\bf 85} 135-196

\bibitem{Bressloff19} Bressloff P C and Kim H 2019 A search-and-capture model of cytoneme-mediated morphogen gradient formation. {\em Phys. Rev. E} {\bf 99} 052401 



\bibitem{Bressloff20} Bressloff P C 2020 Occupation time of a run-and-tumble particle with resetting. {\em Phys. Rev. E} {\bf 102} 042135 

\bibitem{Bressloff22} Bressloff PC 2022  Diffusion-mediated absorption by partially reactive targets: Brownian functionals and generalized propagators. {\em J. Phys. A.} {\bf 55} 205001

\bibitem{Bressloff22a} Bressloff PC 2022 Spectral theory of diffusion in partially absorbing media. {\em Proc. R. Soc. A} {\bf 478} 20220319

\bibitem{Bressloff22b} Bressloff P C 2022  Diffusion-mediated surface reactions and stochastic resetting. {\em J. Phys. A} {\bf 55} 275002 

\bibitem{Bressloff22rtp} Bressloff P C 2022 Encounter-based model of a run-and-tumble particle. {\em J. Stat. mech}. {\bf 113206} (2022).


\bibitem{Demaerel18}  Demaerel T and Maes C 2018 Active processes in one dimension,
{\em Phys. Rev. E} {\bf 97}, 032604  


 \bibitem{Dhar19} Dhar A, Kundu A, Majumdar S N, Sabhapandit S and 
Schehr G 2019 Run-and-tumble particle in one-dimensional confining
potentials: Steady-state, relaxation, and first-passage properties,
{\em Phys. Rev. E} {\bf 99}, 032132

\bibitem{Dogterom93} Dogterom M and Leibler S 1993 Physical aspects of the growth
and regulation of microtubule structures {\em Phys. Rev. Lett.} {\bf 70} 1347-1350  

  \bibitem{Evans18} Evans M R and Majumdar S N 2018 Run and tumble particle under resetting: a renewal approach. {\em J. Phys. A: Math. Theor.} {\bf 51} 475003 (2018).
  
\bibitem{Foerster59} Von Foerster H 1959 Some remarks on changing populations, in {\em The Kinetics of Cellular Proliferation.} edited by F.
Stohlman, Jr. Grune and Stratton, New York  


  \bibitem{Gradenigo19} Gradenigo G and Majumdar S N 2019 A first-order dynamical
transition in the displacement distribution of a driven run-and-tumble
particle {\em J. Stat. Mech.} 053206.



\bibitem{Grebenkov19b} Grebenkov D S 2019  {Spectral theory of imperfect diffusion-controlled
reactions on heterogeneous catalytic surfaces}
{\em J. Chem. Phys.} {\bf 151} 104108

\bibitem{Grebenkov20} Grebenkov D S 2020  {Paradigm shift in diffusion-mediated surface phenomena.} {\em Phys. Rev. Lett.} {\bf 125} 078102  


\bibitem{Grebenkov22} Grebenkov DS. 2022  {An encounter-based approach for restricted diffusion with a gradient drift.}  {\em J. Phys. A.} {\bf 55} 045203 

 \bibitem{Iannelli17} Iannelli M, and Milner F 2017 The basic approach to age-structured population dynamics: models, methods and numerics. {\em Lecture notes on mathematical modelling in the life sciences}. Springer

 \bibitem{Ito65} Ito K and McKean H P 1965 {\em Diffusion Processes and Their Sample Paths} Springer-Verlag,
Berlin

\bibitem{Doussal19} Le Doussal P, Majumdar S N and Schehr G 2019 Non-crossing
run-and-tumble particles on a line {\em Phys. Rev. E} {\bf 100}, 012113

\bibitem{Lee13} Lee C F 2013 Active particles under confinement: aggregation at the wall and gradient formation inside a channel {\em New J. Phys.} {\bf 15} 055007

\bibitem{Majumdar05}  Majumdar S N 2005 Brownian functionals in physics and computer science. {\em Curr. Sci.} {\bf 89}, 2076  

\bibitem{Majumdar20}  Majumdar S N 2020 Toward the full short-time statistics of an active Brownian particle on the plane {\em Phys. Rev. E} {\bf 102} 022113

 \bibitem{Malakar18} Malakar K, Jemseena V, Kundu A, Vijay Kumar, 
Sabhapandit S, Majumdar S N, Redner S and Dhar A 2018 Steady state, relaxation and first-passage properties of a run-and-tumble particle in one-dimension, {\em J. Stat.
Mech.} 043215 

 \bibitem{Martens12}  Martens K, Angelani L, Di Leonardo R and Bocquet L 2012
Probability distributions for the run-and-tumble bacterial dynamics:
An analogy to the Lorentz model {\em Eur. Phys. J. E} {\bf 35}
84
  
 \bibitem{McKean75} McKean H P 1975 {Brownian local time.} {\em Adv. Math.} {\bf 15} 91-111 
 
 \bibitem{McKendrick25} McKendrick A G 1925 Applications of mathematics to medical
problems. {\em Proc. Edinb. Math. Soc. } {\bf 44} 98  
 
 
 \bibitem{Mori20} Mori F, Le Doussal P, Majumdar S N and Schehr G 2020 Universal Survival Probability for a d-Dimensional Run-and-Tumble Particle. {\em Phys. Rev. Lett.} {\bf 124} 090603
 
\bibitem{Mulder12} Mulder B M 2012 Microtubules interacting with a boundary: Mean length and mean first-passage times. {\em Phys. Rev. E} {\bf 86} 011902 


 \bibitem{Newby10} Newby J M and Bressloff P C 2010 Quasi-steady state reduction of molecular-based models of directed intermittent search. {\em Bull. Math. Biol.} {\bf 72} 1840
 
 
\bibitem{Santra20} Santra I, Basu U and Sabhapandit S 2020 Run-and-tumble particles in two-dimensions: Marginal position distributions {\em Phys. Rev. E} {\bf 101} 062120 
 
\bibitem{Santra20a} Santra I, Basu U and Sabhapandit S 2020 Run-and-tumble particles in two dimensions under stochastic resetting conditions {\em J. Stat. Mech.} 113206
 
 \bibitem{Sevilla19}  Sevilla F J, Arzola A V and Cital E P 2019 Stationary superstatistics
distributions of trapped run-and-tumble particles {\em Phys.
Rev. E} {\bf 99}, 012145 

\bibitem{Singh19} Singh P and Kundu A 2019 Generalised ``Arcsine'' laws for run-and-tumble particle in one dimension {\em J. Stat.Mech.} 083205  

\bibitem{Wagner17} Wagner C G, Hagan M F and Baskaran A 2017 Steady-state distributions of ideal active Brownian particles under confinement and forcing {\em J. Stat. Mech.} {\bf 043203} 

\bibitem{Wagner22} Wagner C G, Hagan M F and Baskaran A 2022 Steady states of active Brownian particles interacting with boundaries {\em J. Stat. Mech.} {\bf 013208} 

\bibitem{Zelinski12} Zelinski B, Muller N and Kierfeld J 2022 Dynamics and length distribution of microtubules under force and confinement. {\em Phys. Rev. E} {\bf 86} 041918 



\end{thebibliography}
\end{document}